\documentclass[prc,preprint,showpacs,showkeys,amsmath,amssymb]{revtex4-1}
\pdfoutput=1

\usepackage{graphicx}% Include figure files
\usepackage{dcolumn}% Align table columns on decimal point
\usepackage{bm}% bold math
\usepackage{hyperref}% add hypertext capabilities
\usepackage[all]{hypcap}% make links to figures go to the top of the figure rather than the caption
\usepackage{IEEEtrantools}
\usepackage{todonotes}

\begin{document}

\title{Method for High Accuracy Multiplicity Correlation Measurements}% Force line breaks with \\

\author{K. Gulbrandsen}
\email{gulbrand@nbi.dk}
\affiliation{Niels Bohr Institute, Discovery Center\\Copenhagen, Denmark}

\author{C. S\o{}gaard}
\email{carsten.sogaard@hep.lu.se}
\thanks{Funded by a research grant from the Villum Foundation}
\affiliation{Lund University, Sweden}

\date{\today}

\begin{abstract}
Multiplicity correlation measurements provide insight into the dynamics of high energy collisions.
Models describing these collisions need these correlation measurements to tune the strengths of the underlying QCD processes which influence all observables.
Detectors, however, often possess limited coverage or reduced efficiency that influence correlation measurements in obscure ways.
In this paper, the effects of non-uniform detection acceptance and efficiency on the measurement of multiplicity correlations between two distinct detector regions (termed forward-backward correlations) are derived.
An analysis method with such effects built-in is developed and subsequently verified using different event generators.
The resulting method accounts for acceptance and efficiency in a model independent manner with high accuracy thereby shedding light on the relative contributions of the underlying processes to particle production.
\end{abstract}

%Use showkeys class option if keyword
\maketitle

\section{Introduction}
\label{sec:intro}
The charged particles produced in high energy particle collisions are the result of hard and soft interactions.
The hard processes are well described by perturbative Quantum Chromodynamics while the soft processes, which occur at low momentum and are the bulk of the interactions, are non-perturbative and, therefore, difficult to describe.
This necessitates the use of effective models to characterize these processes.
The models must be verified by (and possibly tuned to) experimental results.
Therefore, characterization of the properties of the distributions of the produced particles is essential for understanding the soft processes involved in the collisions which are also important for understanding the hard processes as they affect the underlying event.
In this paper we focus on the phenomenon called forward-backward particle multiplicity correlations (or forward-backward correlations for short)~\cite{Sjostrand:1987su} to shed light on these soft processes.

Forward-backward correlations measure the correlation strength between the number of particles produced in regions located in opposite hemispheres separated by the plane perpendicular to the beam axis intersecting the collision point.
The regions are typically equidistant (angularly) from the plane perpendicular to the beam axis and probe the forward and backward rapidities where most of the particle production is expected.
This measurement has the advantage that it is mostly influenced by the dynamics of the collision rather than the following hadronization processes~\cite{Hwa:2007sq}.

Models implement the underlying processes in these collisions in different ways.
In Pythia, three main processes exist which affect forward-backward correlations~\cite{Sjostrand:1987su,Wraight:2011ej}.
The first process comprises hard scatterings which generally produce forward-backward correlations limited to small angular separations.
The second process is initial state radiation which is the emittance of gluons at early times during the interaction and generally causes forward-backward correlations with larger angular separations.
The third process is multiple parton interactions which is an effective many-body QCD interaction that causes forward-backward correlations with the largest angular separations.
Various tunes of Pythia arise with different contributions from these processes to particle production~\cite{Skands:2009zm}.
To investigate which tune more accurately describes reality, one needs to either measure forward-backward correlations with large angular separations (where the net effect of the different contributions is most pronounced) or with high accuracy and precision.
Large angular separations are often beyond the design of experiments.
High accuracy and precision require advanced techniques to ensure minimal detector bias and are investigated here.

While different measures exist for characterizing forward-backward correlations, in this paper we focus only on the Pearson correlation factor, which we denote as $b$.
This correlation factor is defined as:
\begin{IEEEeqnarray}{rCl}
b &\equiv& \textrm{Cor}(N_f,N_b)=\frac{\textrm{Cov}(N_f,N_b)}{\sqrt{\textrm{Var}(N_f) \cdot \textrm{Var}(N_b)}} \nonumber \\
&=& \frac{\langle N_fN_b \rangle - \langle N_f \rangle\langle N_b \rangle}
{\sqrt{(\langle N_f^2 \rangle - \langle N_f \rangle^2) \cdot (\langle N_b^2 \rangle - \langle N_b \rangle^2)}}
\label{eq:b}
\end{IEEEeqnarray}
where $N_f$ and $N_b$ are the number of particles produced in the regions in the forward and backward hemispheres, respectively.

One important property of the Pearson correlation factor is that it is a bound quantity.
It can be shown that $-1 \leq b \leq 1$~\cite{soegaardPhD} and does not scale with the multiplicity of the event.
This property arises from the denominator of $b$, which is the square root of the product of the forward and backward multiplicity variances.

\begin{figure}
\includegraphics[width=0.32\textwidth]{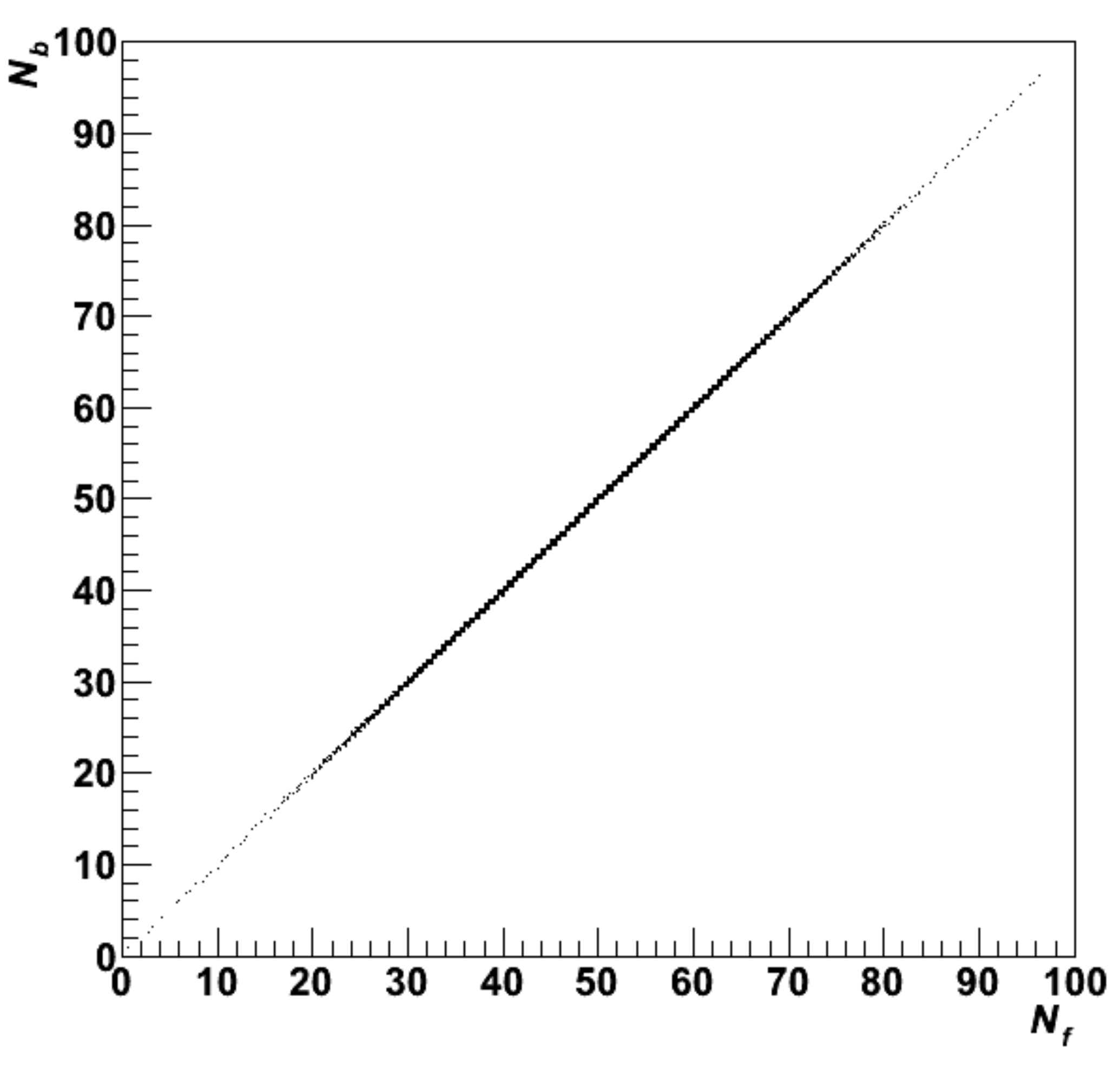}
\includegraphics[width=0.32\textwidth]{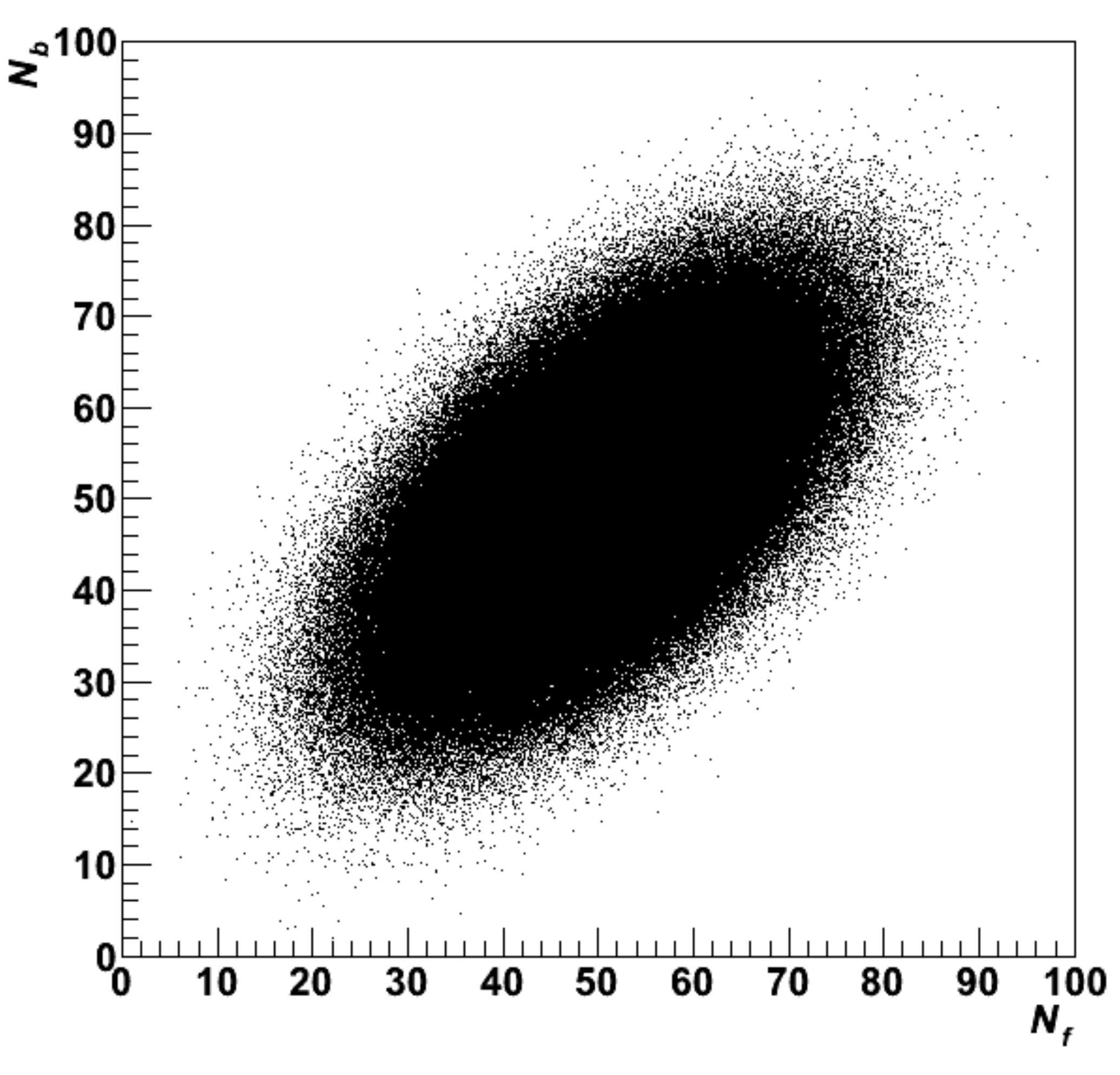}
\includegraphics[width=0.32\textwidth]{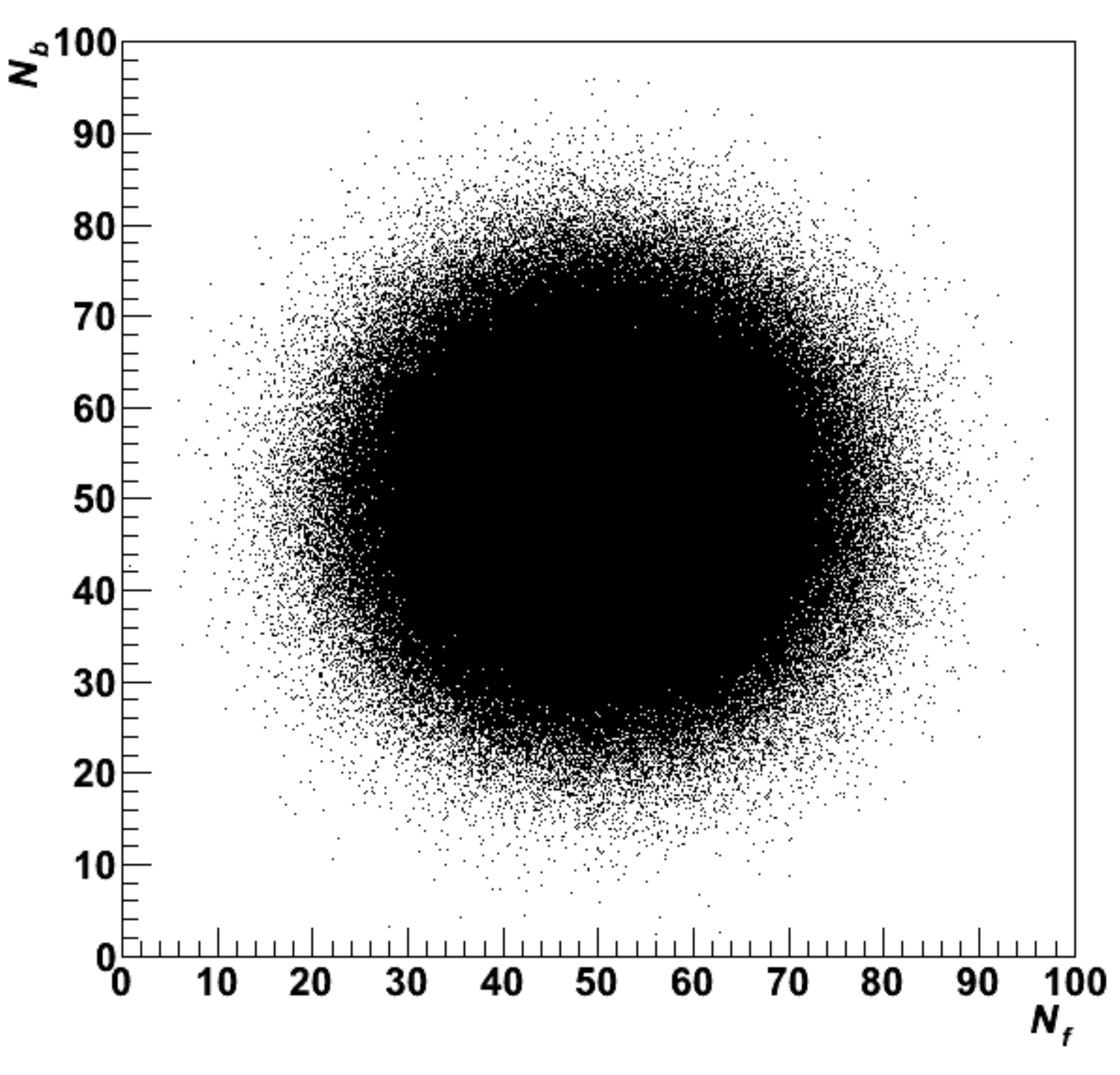}
\caption{The figures depict three sets of forward-backward multiplicity pairs with $b = 1, 0.6, \textrm{and } 0$ from left to right.
The variances (used in the denominator of $b$) are the same for $n_F$ and $n_B$ in all three cases.
This demonstrates that the correlation information is essentially contained in the covariance.}
\label{fig:CorrPlots}
\end{figure}

The correlation factor can be interpreted geometrically as how well the set of number pairs describe a line when plotted on a two dimensional figure.
This is demonstrated in Fig.~\ref{fig:CorrPlots}.
The intersection and the slope of the line are irrelevant to the value of $b$~\cite{soegaardPhD}. This can likewise be demonstrated by the fact that 
\begin{IEEEeqnarray}{rCl}
	\textrm{Cor}(\alpha X + \beta, \gamma Y + \nu) &=& \textrm{Cor}(X,Y), \  \textrm{if } \alpha \gamma > 0
\end{IEEEeqnarray}
where $\alpha$, $\beta$, $\gamma$, and $\nu$ are constants. If $\alpha \gamma < 0$, the correlation factor switches sign.
If the slope in Fig.~\ref{fig:CorrPlots} is negative, the corresponding correlation factor is also negative and the quantities are said to be anti-correlated.

While Eq.~(\ref{eq:b}) shows that only five quantities ($\langle N_f \rangle$, $\langle N_b \rangle$, $\langle N_fN_b \rangle$, $\langle N_f^2 \rangle$ and $\langle N_b^2 \rangle$) are necessary to calculate the correlation factor, the measurement is often not trivial to perform for many detector types.
Any observable will be altered by the environment surrounding the collision in the experiment.
Secondary particle production and partial detector acceptance and inefficiency will influence the measurement.
Directly evaluating this influence in a model independent way is challenging for correlation measurements~\cite{Ravan:2013lwa}.
This is especially evident when evaluating the variances in the denominator of $b$ when partial acceptance exists. The correlation between the measured and not measured regions requires more sophisticated techniques if the gaps in the acceptance are significant.
While the effect of secondary particle production is beyond the scope of this paper (but could be the subject of a subsequent paper), the effect of detector inefficiency and partial detector acceptance is examined.
The influence on the measured correlation strength and a means to account for these effects is provided.
The method is verified through studies using simulations. 
 
\section{Measuring the Correlation Factor}
\label{sec:measuringb}
While forward-backward correlations can be measured in both collider and fixed target experiments, the investigation here is done for collider experiments.
The space surrounding the collision is divided into a forward hemisphere and a backward hemisphere separated by the plane perpendicular to the beam axis intersecting the collision point.
The hemisphere where $\theta<\pi/2$ is usually termed forward, and the hemisphere where $\theta>\pi/2$ is usually termed backward, where the reference direction at $\theta=0$ is defined by the experiment.

Forward-backward multiplicity correlations are usually measured between bins of equal width (in $\eta$) spanning the entire azimuth.
Correlations between bins where only part of the azimuthal angle is taken into account (twist correlations) can also be measured \cite{Wraight:2011ej}.
While these twist correlations are not directly computed in this paper, they require merely a subset of the information necessary to analyze the full azimuth and, therefore, the techniques presented here could be used with minor modifications to measure twist correlations.
The centers of the two bins (in $\eta$) are likewise usually equidistant from $\eta=0$.
In this paper we call such a pair of geometrical regions a {\it forward-backward bin}.

\begin{figure}
	\centering
	\includegraphics[width=0.35\textwidth]{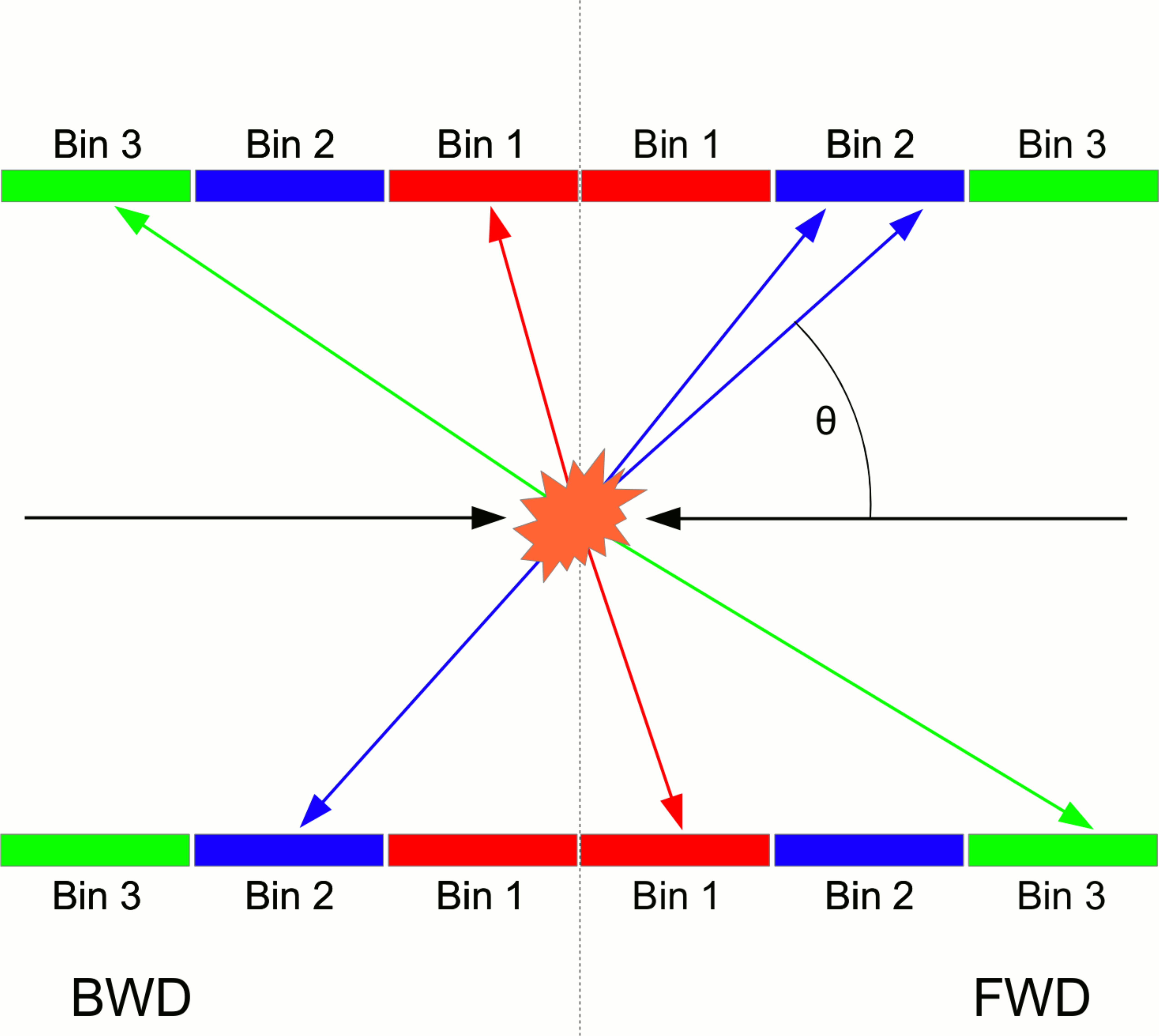}
	\hspace{3ex}
	\includegraphics[width=0.50\textwidth]{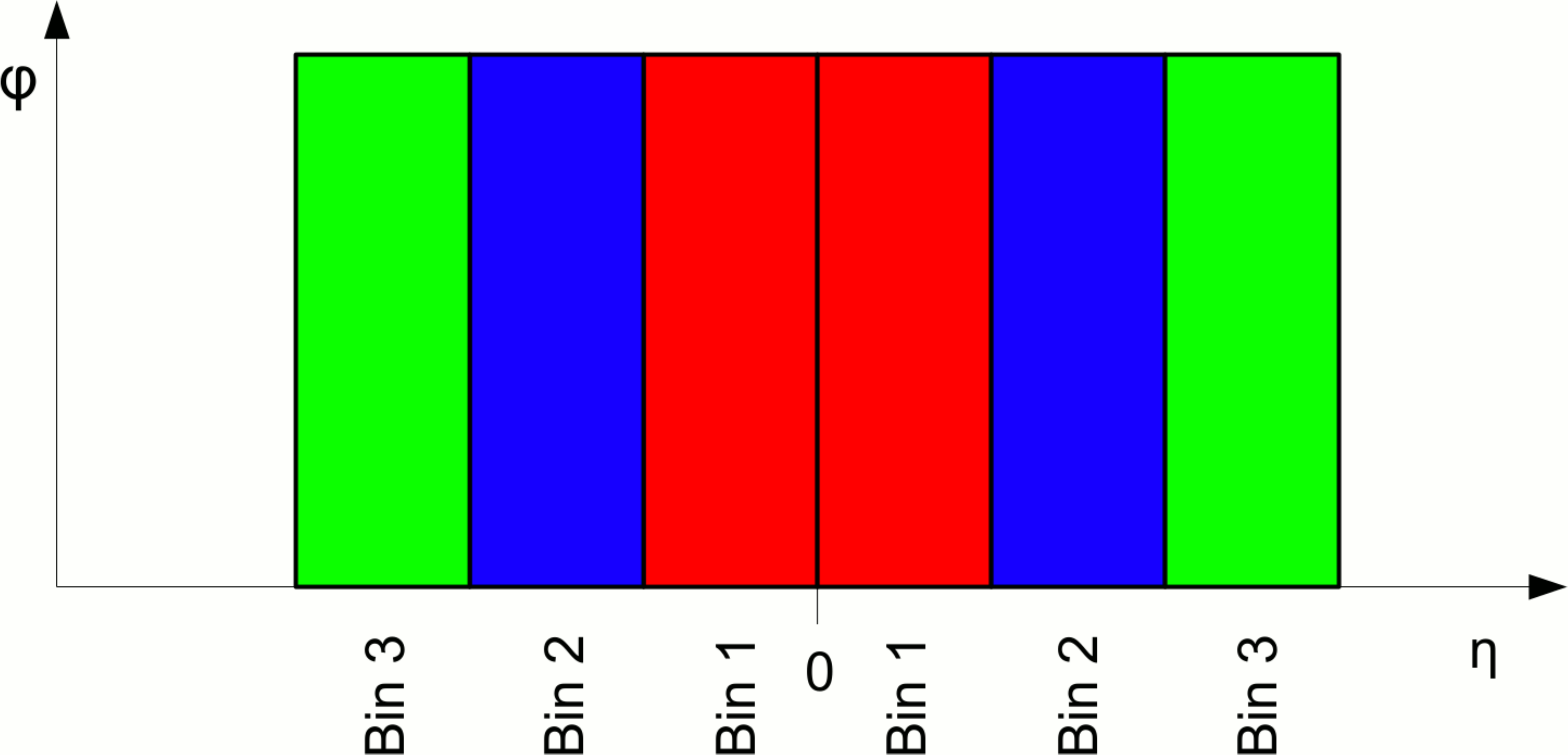}
	\caption{(Color online) Left: The detector is schematically divided into two halves along the line at $\theta=\pi/2$.
Each half then consists of solid angles which span a ``small'' polar angle and the full azimuthal angle.
A forward-backward bin consists of two regions where their centers are equidistant from $\theta=\pi/2$.
The regions are often termed forward when $\theta<\pi/2$ and backward when $\theta>\pi/2$.
Each forward-backward bin is represented by a different color here.
Right: The detector regions have been mapped to a two dimensional figure.
The horizontal axis is the pseudorapidity, which is a function of the polar angle, $\theta$.
Note that the mapping is not to scale and equal size polar angle bins on the left do not correspond to equal size pseudorapidity bins on the right.}
\label{fig:FwdBwdBins}
\end{figure}

The analysis is carried out by determining the number of particles present in each geometrical region event-by-event.
From these particle multiplicities the necessary five quantities are calculated for each event.
These five values are then averaged over all events and the correlation factor is calculated.
Figure~\ref{fig:FwdBwdBins} shows an example of how the forward-backward bins are defined. 

\subsection{The Effect of Efficiency}
\label{sec:derivation}
It is common, either by design or due to malfunction, that detectors do not register all particles impinging on them.
Full hermeticity does not usually exist either.
In both cases, the result is that fewer particles are detected than were actually produced in the collision.
This alters the value of an observable.
First order observables, like the average number of produced particles, can account for this in a straight-forward manner, since the value scales with the efficiency or acceptance.
For higher order observables, the effect of efficiency or acceptance becomes more complex.

To study the effect of efficiency, a statistical approach is taken.
In the case of forward-backward correlations, a joint probability distribution for the produced primary particles, $P^P(N^P_f,N^P_b)$, contains the physics information one wants to measure.
The joint probability distribution is normalized such that
\begin{IEEEeqnarray}{rcl}
	\sum_{N^P_f=0}^\infty \sum_{N^P_b=0}^\infty P^P(N^P_f,N^P_b) &=& 1
	\label{eq:normP}
\end{IEEEeqnarray}
A moment generating function can be defined from this whose derivatives evaluated at $t_f = 0$ and $t_b = 0$ produce all of the desired moments.
\begin{IEEEeqnarray}{rCl}
	\textrm{mgf}^P(t_f,t_b) &\equiv& \sum_{N^P_f=0}^\infty \sum_{N^P_b=0}^\infty P^P(N^P_f,N^P_b) e^{N^P_f t_f + N^P_b t_b}
	\label{eq:mgfP}
\end{IEEEeqnarray}
From the moment generating function, the cumulant generating function is defined as:
\begin{IEEEeqnarray}{rCl}
	\textrm{cgf}^P(t_f,t_b) &\equiv& \ln \left[\textrm{mgf}^P(t_f,t_b)\right]
\end{IEEEeqnarray}
where derivatives of $\textrm{cgf}^P$ evaluated at $t_f = 0$ and $t_b = 0$ produce the quantities desired to compute the correlation factor (and many more cumulants with further derivatives).
For the purpose of this paper, the first two cumulants (the mean and the covariance) are important.
\begin{IEEEeqnarray}{rCl}
	\frac{\partial\textrm{cgf}^P}{\partial t_r}(0,0) &\equiv& \textrm{cgf}^P_r(0,0) = \langle N_r \rangle, \  \textrm{where } r = f \textrm{ or } b
	\label{eq:firstcumulant} \\
	 \frac{\partial^2\textrm{cgf}^P}{\partial t_{r_1} \partial t_{r_2}}(0,0) &\equiv& \textrm{cgf}^P_{r_1 r_2}(0,0) = \textrm{Cov}(N^P_{r_1}, N^P_{r_2}), \  \textrm{where } r_1,r_2 = f \textrm{ or } b
	 \label{eq:secondcumulant}
\end{IEEEeqnarray}
In Eq.~(\ref{eq:firstcumulant}), $r$ stands for a ``region'' that could be forward or backward.
In Eq.~(\ref{eq:secondcumulant}), $r_1$ and $r_2$ stand for ``region 1'' and ``region 2'', respectively, and can independently be forward or backward.
In the case where $r_1 = r_2 \textrm{ (} = r \textrm{)}$, the covariance becomes the variance, such that
\begin{IEEEeqnarray}{rCl}
	\textrm{Cov}(N^P_r, N^P_r) &=& \textrm{Var}(N^P_r)
\end{IEEEeqnarray}

We now consider the case where a uniform detection efficiency exists over the whole forward and backward regions ($\varepsilon_f$ and $\varepsilon_b$ respectively).
Perfect detection efficiency is defined to have a value of 1 while a completely dead region would have a value of 0.
The restriction of uniformity is not realistic, but is instructive for an initial investigation where the efficiency will be taken to be the average detection efficiency in the region.
Equation~(\ref{eq:normP}) is then modified as follows to account these efficiencies in the forward and backward regions.
\begin{IEEEeqnarray}{rCl}
	\sum_{N^P_f=0}^\infty \sum_{N^P_b=0}^\infty P^P(N^P_f,N^P_b) \left(\varepsilon_f+(1-\varepsilon_f)\right)^{N^P_f} \left(\varepsilon_b+(1-\varepsilon_b)\right)^{N^P_b} &=& 1
	\label{eq:normE}
\end{IEEEeqnarray}

One can now arrive at the moment generating function for the detected particles ($\textrm{mgf}^D$).
Since one particle is detected with the probability $\varepsilon_r$, one applies the term $e^{1 \cdot t_r}$ to the $\varepsilon_r$ terms.
Likewise, one applies $e^{0 \cdot t_r}=1$ to the $(1-\varepsilon_r)$ terms, since no particle is detected with this probability.
The resulting term, $\varepsilon_r e^{t_r}+(1-\varepsilon_r)$, is actually the moment generating function for a specific particle to be found in the region with probability $\varepsilon_r$, which we term $\textrm{mgf}^E (t_r;\varepsilon_r)$.
The corresponding cumulant generating function is then $\textrm{cgf}^E (t_r;\varepsilon_r) \equiv \ln \left[ \textrm{mgf}^E (t_r;\varepsilon_r) \right]$.
The moment generating function for the distribution of detected particles then becomes:
\begin{IEEEeqnarray}{rCl}
	\textrm{mgf}^D(t_f, t_b) &=& \sum_{N^P_f=0}^\infty \sum_{N^P_b=0}^\infty P^P(N^P_f,N^P_b)\left[\varepsilon_f e^{t_f}+(1-\varepsilon_f)\right]^{N^P_f} \left[\varepsilon_b e^{t_b}+(1-\varepsilon_b)\right]^{N^P_b} \nonumber \\
	&=& \sum_{N^P_f=0}^\infty \sum_{N^P_b=0}^\infty P^P(N^P_f,N^P_b) e^{N^P_f \textrm{cgf}^E(t_f;\varepsilon_f) + N^P_b \textrm{cgf}^E(t_b;\varepsilon_b)}
	\label{eq:mgfD}
\end{IEEEeqnarray}
Comparing Eq.~(\ref{eq:mgfD}) to Eq.~(\ref{eq:mgfP}) shows that the effect of detection efficiency is merely a substitution of the variables in the moment generating function of primary particles, namely $t_r \rightarrow \textrm{cgf}^E (t_r;\varepsilon_r)$.
The final equation relating the cumulant generating function of detected particles to the cumulant generating function of primary particles is then found by:
\begin{IEEEeqnarray}{rCl}
	\textrm{mgf}^D (t_f, t_b) &=& \textrm{mgf}^P \left(\textrm{cgf}^E (t_f;\varepsilon_f), \textrm{cgf}^E (t_b;\varepsilon_b)\right) \Rightarrow \nonumber \\
	\textrm{cgf}^D (t_f, t_b) &=& \textrm{cgf}^P \left(\textrm{cgf}^E (t_f;\varepsilon_f), \textrm{cgf}^E (t_b;\varepsilon_b)\right)
	\label{eq:cfg_detected}
\end{IEEEeqnarray}
One should note that Eq.~(\ref{eq:cfg_detected}) can be generalized to allow one to evaluate the effect of acceptance or efficiency on any order correlation.
The $n$-region equivalent of Eq.~(\ref{eq:cfg_detected}) is:
\begin{IEEEeqnarray}{rCl}
	\textrm{mgf}^D (t_1, \cdots, t_n) &=& \textrm{mgf}^P \left(\textrm{cgf}^E (t_1;\varepsilon_1), \cdots, \textrm{cgf}^E (t_n;\varepsilon_n)\right) \Rightarrow \nonumber \\
	\textrm{cgf}^D (t_1, \cdots, t_n) &=& \textrm{cgf}^P \left(\textrm{cgf}^E (t_1;\varepsilon_1), \cdots, \textrm{cgf}^E (t_n;\varepsilon_n)\right)
	\label{eq:cfg_detected_general}
\end{IEEEeqnarray}
Derivatives of Eq.~(\ref{eq:cfg_detected_general}) evaluated at $t_1, \cdots, t_n = 0$ reveal the effect of acceptance or efficiency on the desired moment or cumulant relative to the moments or cumulants of the primary distribution.
One could use this information (as will be done here for the variance and covariance) to account for these effects in higher order correlations.

The cumulants of the distribution of detected particles can now be calculated by differentiating Eq.~(\ref{eq:cfg_detected}) and evaluating the results at $t_f=0$ and $t_b=0$.
The first derivative gives the average number of found particles in a region.
\begin{IEEEeqnarray}{rCl}
	\left.\frac{\partial \textrm{cgf}^D}{\partial t_r}\right|_{t_f,t_b=0} &=& \textrm{cgf}^P_r \left(\textrm{cgf}^E (0;\varepsilon_f), \textrm{cgf}^E (0;\varepsilon_b)\right) \cdot \left. \frac{\textrm{d} \left[\textrm{cgf}^E (t_r;\varepsilon_r)\right]}{\textrm{d} t_r} \right|_{t_r=0} \nonumber \\
        &=& \textrm{cgf}^P_r \left(0, 0\right) \cdot \left. \frac{\varepsilon_r e^{t_r}}{\varepsilon_r e^{t_r} + (1 - \varepsilon_r)} \right|_{t_r=0} \Rightarrow \nonumber \\
	\langle N^D_r \rangle &=& \langle N^P_r \rangle \cdot \varepsilon_r
        \label{eq:DetectedMean}
\end{IEEEeqnarray}

The result in Eq.~(\ref{eq:DetectedMean}) is expected, since it is intuitive that the mean value of the distribution scales with the probability that any given particle is detected.
The variances or the covariance (given by the second derivative), however, yield a more complicated result.
\begin{IEEEeqnarray}{rCl}
  	\left. \frac{\partial^2 \textrm{cgf}^D}{\partial t_{r_1} \partial t_{r_2}} \right|_{t_f,t_b=0} &=&
	\textrm{cgf}^P_{r_1 r_2} \left(\textrm{cgf}^E(0;\varepsilon_f), \textrm{cgf}^E(0;\varepsilon_b)\right)
	\cdot \left. \frac{\textrm{d}\left[\textrm{cgf}^E \left(t_{r_1};\varepsilon_{r_1}\right) \right]}{\textrm{d} t_{r_1}} \right|_{t_{r_1}=0}
	\cdot \left. \frac{\textrm{d}\left[\textrm{cgf}^E \left(t_{r_2};\varepsilon_{r_2}\right) \right]}{\textrm{d} t_{r_2}} \right|_{t_{r_2}=0} \nonumber \\
	&& + \delta_{r_1 r_2} \cdot \textrm{cgf}^P_{r_1} \left(\textrm{cgf}^E(0;\varepsilon_f), \textrm{cgf}^E(0;\varepsilon_b)\right)
	\cdot \left.\frac{\textrm{d}^2\left[\textrm{cgf}^E(t_{r_1};\varepsilon_{r_1})\right]}{\textrm{d} t^2_{r_1}}\right|_{t_{r_1}=0} \nonumber \\
	&=& \textrm{cgf}^P_{r_1 r_2} \left(0,0\right)
	\cdot \left. \frac{\varepsilon_{r_1} e^{t_{r_1}}}{\varepsilon_{r_1} e^{t_{r_1}} + (1 - \varepsilon_{r_1})} \right|_{t_{r_1}=0}
	\cdot \left. \frac{\varepsilon_{r_2} e^{t_{r_2}}}{\varepsilon_{r_2} e^{t_{r_2}} + (1 - \varepsilon_{r_2})} \right|_{t_{r_2}=0} \nonumber \\
	&& + \delta_{r_1 r_2} \cdot \textrm{cgf}^P_{r_1} \left(0,0\right)
	 \cdot \left. \frac{\left[ \varepsilon_{r_1} e^{t_{r_1}} + \left(1 - \varepsilon_{r_1}\right) \right] \cdot \varepsilon_{r_1} e^{t_{r_1}} - \left( \varepsilon_{r_1} e^{t_{r_1}} \right)^2}{\left[\varepsilon_{r_1} e^{t_{r_1}} + \left(1 - \varepsilon_{r_1}\right)\right]^2} \right|_{t_{r_1}=0} \Rightarrow \nonumber \\
	\textrm{Cov}(N^D_{r_1},N^D_{r_2}) &=&
	\textrm{Cov}(N^P_{r_1},N^P_{r_2}) \cdot \varepsilon_{r_1} \varepsilon_{r_2}
	+ \delta_{r_1 r_2} \cdot \langle N^P_{r_1} \rangle \cdot \varepsilon_{r_1} \left( 1 - \varepsilon_{r_1} \right)
	\label{eq:DetectedCovVar}
\end{IEEEeqnarray}
This result shows that a special case exists for the variance where the differentiation is performed twice with respect to the same variable and the Kronecker delta ($\delta_{r_1 r_2}$) evaluates to 1.
The final expressions for the covariance and the variances of the distribution of detected particles are:
\begin{IEEEeqnarray}{rCl}
	\textrm{Cov}(N^D_f,N^D_b) &=& \textrm{Cov}(N^P_f,N^P_b) \cdot \varepsilon_f \varepsilon_b \label{eq:DetectedCov}\\
	\textrm{Var}(N^D_r) &=& \textrm{Var}(N^P_r) \cdot \varepsilon^2_r + \langle N^P_r \rangle \cdot \varepsilon_r \left( 1 - \varepsilon_r \right) \label{eq:DetectedVar}
\end{IEEEeqnarray}
Equation~(\ref{eq:DetectedVar}) shows that the detected variance has an additional dependence, beyond the variance of the primary produced particles and the efficiency, on the mean number of particles produced in the region, which the covariance does not possess.

Equations~(\ref{eq:DetectedMean}), (\ref{eq:DetectedCov}), and (\ref{eq:DetectedVar}) can be inverted to obtain the cumulants of the distribution of the primary particles from the detected quantities:
\begin{IEEEeqnarray}{rCl}
	\langle N^P_r \rangle &=& \frac{\langle N^D_r \rangle}{\varepsilon_r} \label{eq:CorrectedMean} \\
	\textrm{Cov}(N^P_f,N^P_b) &=& \frac{\textrm{Cov}(N^D_f,N^D_b)}{\varepsilon_f \varepsilon_b} \label{eq:CorrectedCov} \\
	\textrm{Var}(N^P_r) &=& \frac{\textrm{Var}(N^D_r) - \langle N^D_r \rangle \cdot \left( 1 - \varepsilon_r \right)}{\varepsilon^2_r} \label{eq:CorrectedVar}
\end{IEEEeqnarray}
From these expressions, the correlation factor in the case of an imperfect detector (with an efficiency less than 1) is derived as:
\begin{IEEEeqnarray}{rCl}
	b &=& \frac{\textrm{Cov}(N^P_f,N^P_b)}
	{\sqrt{\textrm{Var}(N^P_f) \cdot
	\textrm{Var}(N^P_b)}} \nonumber \\
	&=& \frac{\frac{\textrm{Cov}(N^D_f,N^D_b)}{\varepsilon_f \varepsilon_b}}
        {\sqrt{\frac{\textrm{Var}(N^D_f) - \langle N^D_f \rangle \cdot \left( 1 - \varepsilon_f \right)}{\varepsilon^2_f}}
        \sqrt{\frac{\textrm{Var}(N^D_b) - \langle N^D_b \rangle \cdot \left( 1 - \varepsilon_b \right)}{\varepsilon^2_b}}} \nonumber \\
	&=& \frac{\textrm{Cov}(N^D_f,N^D_b)}
	{\sqrt{\textrm{Var}(N^D_f) - \langle N^D_f \rangle \cdot \left( 1 - \varepsilon_f \right)}
	\sqrt{\textrm{Var}(N^D_b) - \langle N^D_b \rangle \cdot \left( 1 - \varepsilon_b \right)}}
\label{eq:BasicAccCorrection}
\end{IEEEeqnarray}
While the overall multiplicative efficiency factors in the covariance and variance terms cancel when calculating the correlation factor, Eq.~(\ref{eq:BasicAccCorrection}) shows that the additive terms, proportional to the mean number of particles detected in the region, remain and must be evaluated when an inefficiency exists.

The result in Eq.~(\ref{eq:BasicAccCorrection}) assumes that the detection efficiencies, $\varepsilon_f$ and $\varepsilon_b$, are the same for all particles in their respective regions.
When the efficiency varies little or not at all over the region, this assumption is valid.
However, variations in the efficiency of the region will affect a correlation measurement.
The most extreme variation exists when a fraction of the region has no detection efficiency and the rest has perfect detection efficiency, which could be the case when the acceptance of the detector does not cover the whole region (in azimuth for instance).
Additionally, a non-uniform distribution of particles (termed ``event shape'') in the region will affect the measurement when the efficiency varies.
In this case, when the particle multiplicity density is higher in the active region relative to the dead region, the effective efficiency is higher.
The opposite is true when the particle multiplicity density is lower in the active region relative to the dead region.
The net effect does not necessarily cancel out on average over many events when performing correlation measurements.
The effect of efficiency variations and event shape is analyzed in section~\ref{sec:eventshape} using the same framework developed so far and the effects they have on the correlation measurements are examined in section~\ref{sec:verification}.

\subsection{Accounting for Azimuthal Event Shape}
\label{sec:eventshape}
The effect of the event shape (in the presence of an inefficiency) can be reduced if one can select regions of the detector where the particle multiplicity density gradient is small or the efficiency is constant over the region.
This generally occurs when smaller regions of the detector are used.
We first consider the case where, event-by-event, a non-uniform azimuthal event shape exists for the produced particles, which is, however, uniform on average over many events.
The solution is then to segment the $\eta$ regions, studied in the section~\ref{sec:derivation}, additionally into $\varphi$ segments.
The particle multiplicity of these sub-regions will be denoted with an extra subscript (for example, $N^P_{f,1}$ for the primary multiplicity in the first $\varphi$ segment of the forward region), where the second subscript is a value between $1$ and $m_\varphi$ (the number of $\varphi$ segments).
The results in Eqs.~(\ref{eq:DetectedMean}) and (\ref{eq:DetectedCovVar}) have no assumption about the type of segmentation and are, therefore, also true for these sub-regions.
The generalization to these sub-regions is
\begin{IEEEeqnarray}{rCl}
	\langle N^D_{r,i_\varphi} \rangle &=& \langle N^P_{r,i_\varphi} \rangle \cdot \varepsilon_{r,i_\varphi} \label{eq:DetectedMeanSeg} \\
	\textrm{Cov}(N^D_{r_1,i_\varphi},N^D_{r_2,j_\varphi}) &=&
	\textrm{Cov}(N^P_{r_1,i_\varphi},N^P_{r_2,j_\varphi}) \cdot \varepsilon_{r_1,i_\varphi} \varepsilon_{r_2,j_\varphi} \nonumber\\
	&& + \delta_{r_1 r_2} \cdot \delta_{i_\varphi j_\varphi} \cdot \langle N^P_{r_1,i_\varphi} \rangle \cdot \varepsilon_{r_1,i_\varphi} \left( 1 - \varepsilon_{r_1,i_\varphi} \right)
	\label{eq:DetectedCovVarSeg}
\end{IEEEeqnarray}
where $1 \leq i_\varphi \leq m_\varphi$ and $1 \leq j_\varphi \leq m_\varphi$.

The relationship of the mean and covariance of the sub-regions (for primary particles) can be trivially derived.
For the mean, this is
\begin{IEEEeqnarray}{rCl}
	\langle N^P_r \rangle = \langle \sum_{i_\varphi=1}^{m_\varphi} N^P_{r,i_\varphi} \rangle = \sum_{i_\varphi=1}^{m_\varphi} \langle N^P_{r,i_\varphi} \rangle
	\label{eq:SumOfMeans}
\end{IEEEeqnarray}
which is the expected sum of the means of the sub-regions.
For the covariance, this is
\begin{IEEEeqnarray}{rCl}
	\textrm{Cov}(N^P_{r_1},N^P_{r_2}) &=& \langle N^P_{r_1} N^P_{r_2} \rangle - \langle N^P_{r_1} \rangle \langle N^P_{r_2} \rangle \nonumber \\
	&=& \langle \sum_{i_\varphi=1}^{m_\varphi} N^P_{r_1,i_\varphi} \sum_{j_\varphi=1}^{m_\varphi} N^P_{r_2,j_\varphi} \rangle - \langle \sum_{i_\varphi=1}^{m_\varphi} N^P_{r_1,i_\varphi} \rangle \langle \sum_{j_\varphi=1}^{m_\varphi} N^P_{r_2,j_\varphi} \rangle \nonumber \\
	&=& \sum_{i_\varphi=1}^{m_\varphi} \sum_{j_\varphi=1}^{m_\varphi} \langle N^P_{r_1,i_\varphi} N^P_{r_2,j_\varphi} \rangle - \sum_{i_\varphi=1}^{m_\varphi} \sum_{j_\varphi=1}^{m_\varphi} \langle N^P_{r_1,i_\varphi} \rangle \langle N^P_{r_2,j_\varphi} \rangle \nonumber \\
	&=& \sum_{i_\varphi=1}^{m_\varphi} \sum_{j_\varphi=1}^{m_\varphi} \left( \langle N^P_{r_1,i_\varphi} N^P_{r_2,j_\varphi} \rangle - \langle N^P_{r_1,i_\varphi} \rangle \langle N^P_{r_2,j_\varphi} \rangle \right) \nonumber \\
	&=& \sum_{i_\varphi=1}^{m_\varphi} \sum_{j_\varphi=1}^{m_\varphi} \textrm{Cov}(N^P_{r_1,i_\varphi},N^P_{r_2,j_\varphi})
	\label{eq:SumOfCovariances}
\end{IEEEeqnarray}
which is the sum of the covariances of each sub-region to every other sub-region.
One should note that Eqs.~(\ref{eq:SumOfMeans}) and (\ref{eq:SumOfCovariances}) apply also to the detected means and covariances.

To account for acceptance and efficiency, rotational invariance is exploited.
One would expect, for example, that the mean number of primary particles produced at a certain pseudorapidity and at a certain azimuthal angle would be independent of the azimuthal angle (and only dependent on the azimuthal range of the measurement). 
To use this in practice, we will impose the restriction that each $\eta$ region is {\it equally} divided into $m_\varphi$ azimuthal segments that span $2 \pi / m_\varphi$.
With this restriction, many of the measurements are redundant.
For the mean number of primary particles, this means that the value at each angle can be replaced by the average.
\begin{IEEEeqnarray}{rCl}
	\langle N^P_{r,i_\varphi} \rangle = \frac{\sum_{j_\varphi=1}^{m_\varphi}\langle N^P_{r,j_\varphi}\rangle}{m_\varphi}, \textrm{ where } 1 \leq i_\varphi \leq m_\varphi \label{eq:meanrotinv}
\end{IEEEeqnarray}
Using Eqs.~(\ref{eq:DetectedMeanSeg}) and (\ref{eq:meanrotinv}) one can derive the (expected) relationship between the mean number of primary particles and the detected quantities.
\begin{IEEEeqnarray}{rCl}
	\langle N^D_{r} \rangle &=&
		\sum_{i_\varphi=1}^{m_\varphi}\langle N^D_{r,i_\varphi}\rangle =
		\sum_{i_\varphi=1}^{m_\varphi}\langle N^P_{r,i_\varphi}\rangle\varepsilon_{r,i_\varphi} =
		\sum_{i_\varphi=1}^{m_\varphi}\left(\frac{\sum_{j_\varphi=1}^{m_\varphi}\langle N^P_{r,j_\varphi}\rangle}{m_\varphi}\right)\varepsilon_{r,i_\varphi} \nonumber \\
	&=&	\left(\frac{\sum_{j_\varphi=1}^{m_\varphi}\langle N^P_{r,j_\varphi}\rangle}{m_\varphi}\right) \cdot \sum_{i_\varphi=1}^{m_\varphi}\varepsilon_{r,i_\varphi} =
		\frac{\langle N^P_r\rangle}{m_\varphi} \cdot \sum_{i_\varphi=1}^{m_\varphi}\varepsilon_{r,i_\varphi} \nonumber \\
	\Rightarrow \langle N^P_r \rangle &=& 
		m_\varphi \cdot \frac{\sum_{i_\varphi=1}^{m_\varphi}\langle N^D_{r,i_\varphi}\rangle}{\sum_{i_\varphi=1}^{m_\varphi}\varepsilon_{r,i_\varphi}}
	\label{eq:FinalMeanFormula}
\end{IEEEeqnarray}

\begin{figure}
\begin{minipage}{0.32\textwidth}
\includegraphics[width=0.9\textwidth]{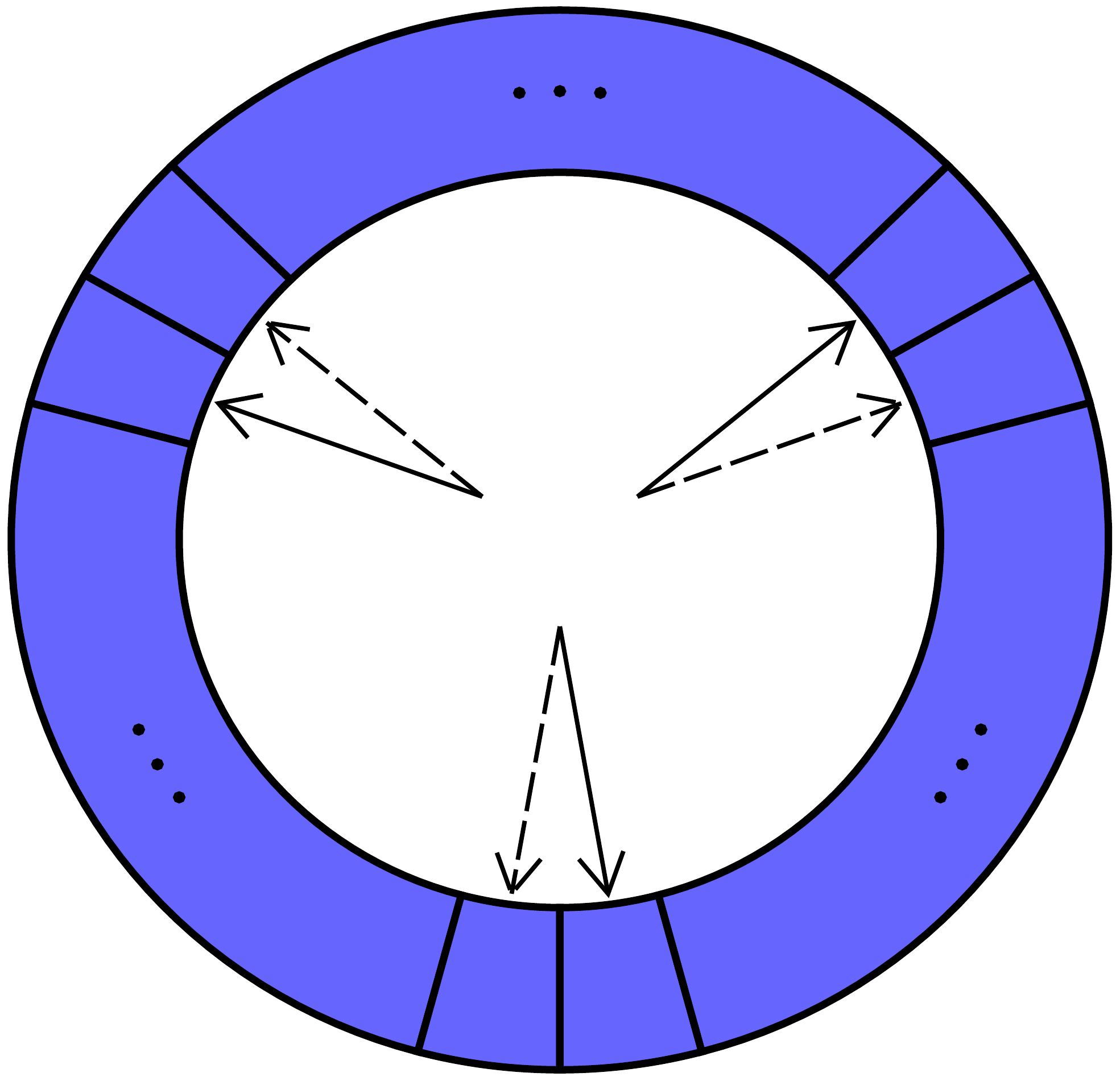}
\end{minipage}
\begin{minipage}{0.32\textwidth}
\includegraphics[width=0.9\textwidth]{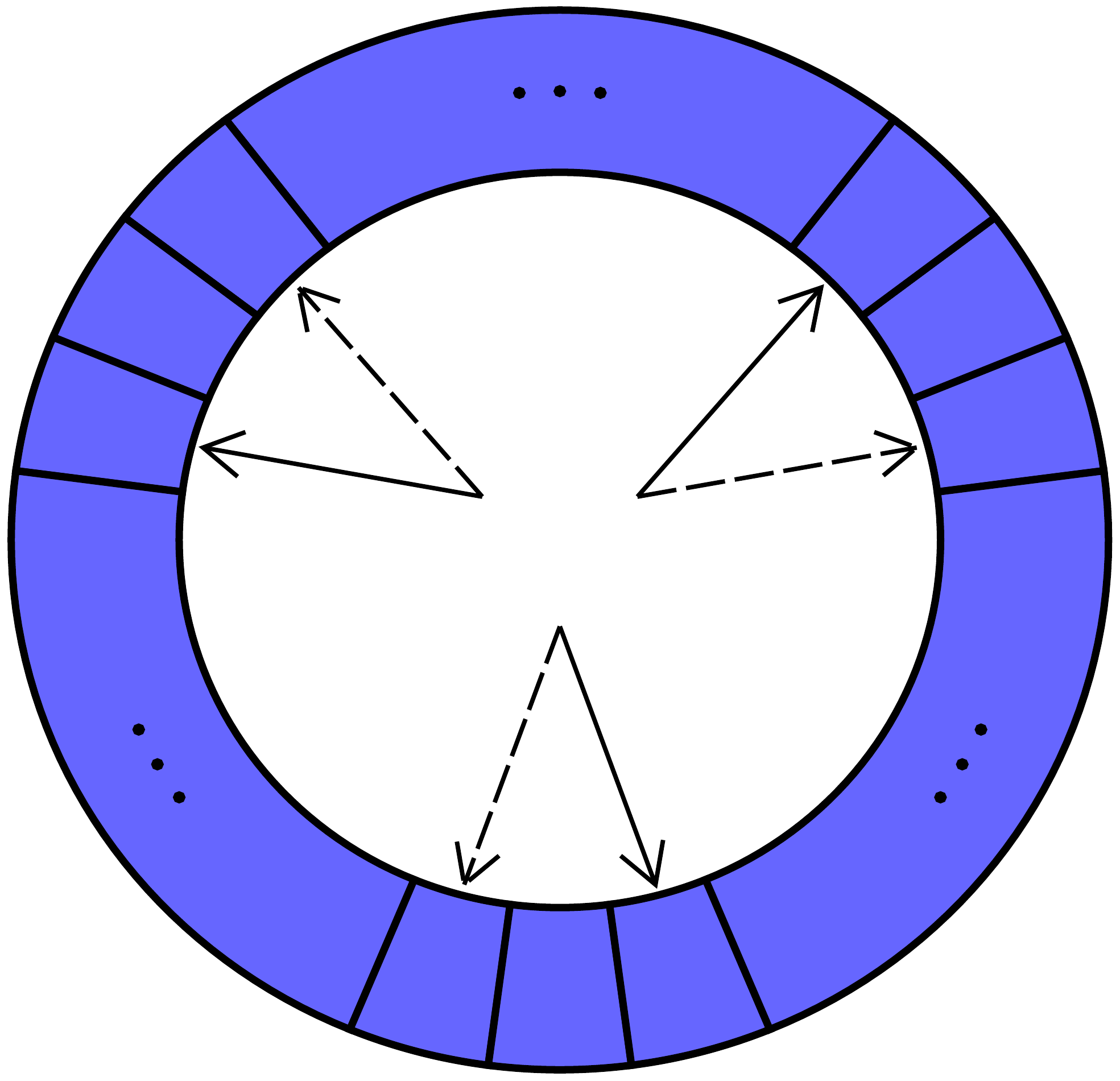}
\end{minipage}
\begin{minipage}{0.32\textwidth}
\includegraphics[width=0.9\textwidth]{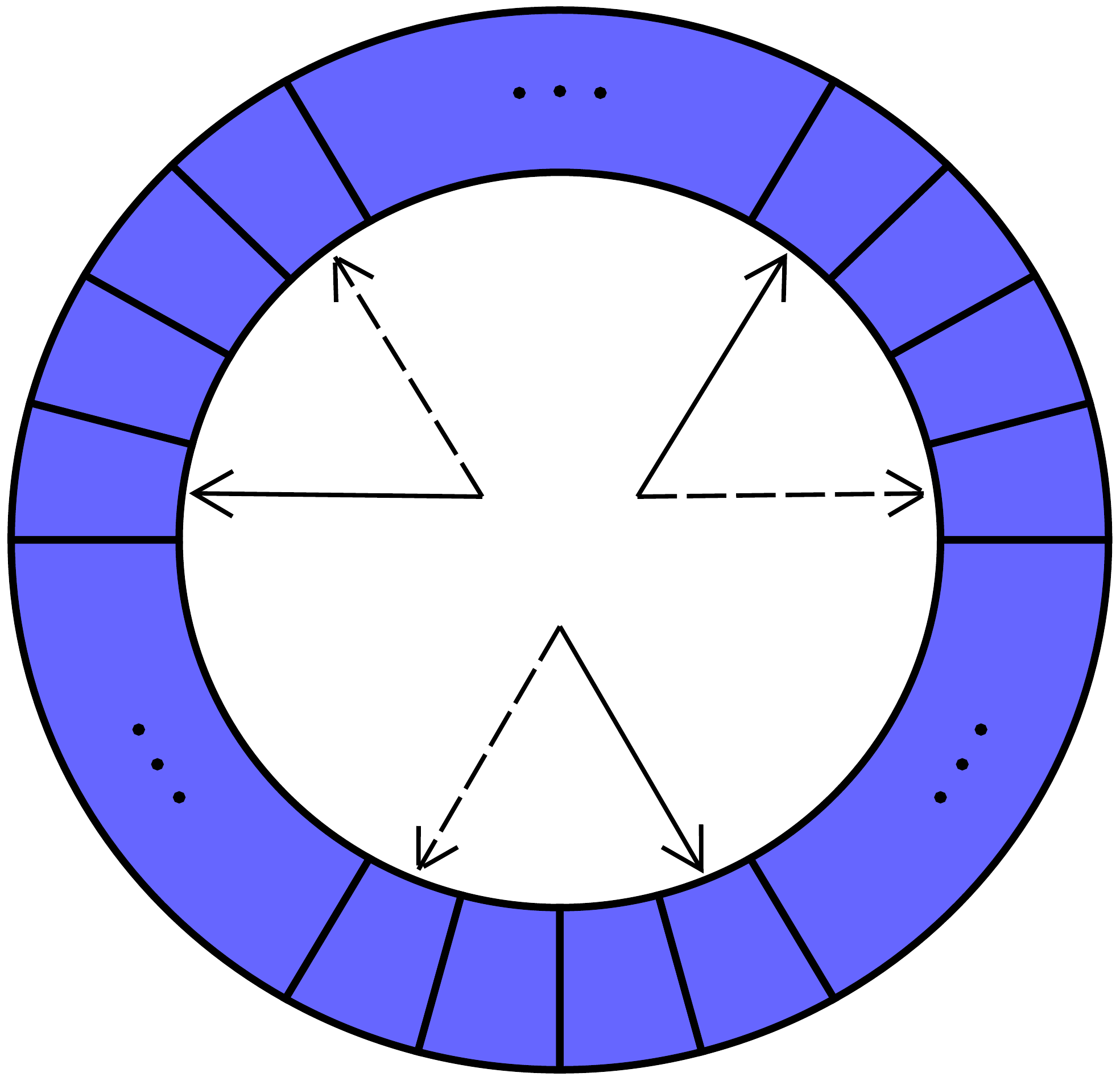}
\end{minipage}
\caption{(Color online) The figures depict sets of correlations between different $\varphi$ regions.
Note that the solid arrow and dashed arrow can (in general) point to different $\eta$ regions.
From left to right, the plots show the correlations between regions shifted by 1, 2, and 3 $\varphi$ segments.
In general this shift can be any value between 0 and the number of $\varphi$ segments minus 1.
For each shift, the correlation between the regions should be the same (for primary particles) independent of the average $\varphi$ angle of the correlated regions.
This produces redundant measures of the same correlation.
A set of redundant measures is called an ``invariant twist''.}
\label{fig:InvariantTwists}
\end{figure}

Equation~(\ref{eq:FinalMeanFormula}) is simple because all quantities in the sum are equivalent (due to rotational invariance).
Rotational invariance can be applied to the expression for the covariance where one expects the covariance between any two segments with equal $\varphi$ displacement to be equivalent (shown in Fig.~\ref{fig:InvariantTwists}).
To do this, Eq.~(\ref{eq:SumOfCovariances}) must be rewritten to group these quantities.
\begin{IEEEeqnarray}{rCl}
	\textrm{Cov}(N^P_{r_1},N^P_{r_2}) &=& \sum_{i_\varphi=1}^{m_\varphi} \textrm{Cov}(N^P_{r_1,i_\varphi},N^P_{r_2,i_\varphi}) \nonumber \\
	&& + \sum_{s=1}^{m_\varphi-1} \left\{
	\sum_{i_\varphi=1}^{m_\varphi-s} \textrm{Cov}(N^P_{r_1,i_\varphi},N^P_{r_2,i_\varphi+s})
	\vphantom{+ \sum_{i_\varphi=1}^{s} \textrm{Cov}(N^P_{r_1,m_\varphi+i_\varphi-s},N^P_{r_2,i_\varphi})}
	\right. \nonumber \\
	&& \hphantom{+ \sum_{s=1}^{m_\varphi-1} \left\{
	\vphantom{\sum_{i_\varphi=1}^{m_\varphi-s} \textrm{Cov}(N^P_{r_1,i_\varphi},N^P_{r_2,i_\varphi+s})
	+ \sum_{i_\varphi=1}^{s} \textrm{Cov}(N^P_{r_1,m_\varphi+i_\varphi-s},N^P_{r_2,i_\varphi})}
	\right.}
	\left.
	\vphantom{\sum_{i_\varphi=1}^{m_\varphi-s} \textrm{Cov}(N^P_{r_1,i_\varphi},N^P_{r_2,i_\varphi+s})}
	+ \sum_{i_\varphi=1}^{s} \textrm{Cov}(N^P_{r_1,m_\varphi+i_\varphi-s},N^P_{r_2,i_\varphi})
	\right\}
	\label{eq:SumOfCovariancesTwists}
\end{IEEEeqnarray}
The first sum in Eq.~(\ref{eq:SumOfCovariancesTwists}) correlates all regions with the same $\varphi$.
The terms within the braces in Eq.~(\ref{eq:SumOfCovariancesTwists}) correlate regions shifted by $s$ segments in $\varphi$ (these correspond to twist correlations).
Every term in the first sum must be the same (on average) by rotational invariance as well as every term within the braces (for each value of $s$).
Each of these terms can be analyzed individually to see how they relate to the detected quantities.

We first analyze the terms inside the braces of Eq.~(\ref{eq:SumOfCovariancesTwists}), but investigate the result as if they were detected quantities instead.
This yields the following result, if rotational invariance is applied for each twisted quantity.
\begin{IEEEeqnarray}{l}
	\sum_{i_\varphi=1}^{m_\varphi-s} \textrm{Cov}(N^D_{r_1,i_\varphi},N^D_{r_2,i_\varphi+s})
	+ \sum_{i_\varphi=1}^{s} \textrm{Cov}(N^D_{r_1,m_\varphi+i_\varphi-s},N^D_{r_2,i_\varphi}) = \nonumber \\
	\phantom{\sum_{i_\varphi=1}^{m_\varphi-s} =} \sum_{i_\varphi=1}^{m_\varphi-s} \textrm{Cov}(N^P_{r_1,i_\varphi},N^P_{r_2,i_\varphi+s}) \cdot \varepsilon_{r_1,i_\varphi} \varepsilon_{r_2,i_\varphi+s} \nonumber \\
	\phantom{\sum_{i_\varphi=1}^{m_\varphi-s} =} + \sum_{i_\varphi=1}^{s} \textrm{Cov}(N^P_{r_1,m_\varphi+i_\varphi-s},N^P_{r_2,i_\varphi}) \cdot \varepsilon_{r_1,m_\varphi+i_\varphi-s} \varepsilon_{r_2,i_\varphi} \nonumber \\
	\phantom{\sum_{i_\varphi=1}^{m_\varphi-s}} = \frac{1}{m_\varphi} \left( \sum_{i_\varphi=1}^{m_\varphi-s} \textrm{Cov}(N^P_{r_1,i_\varphi},N^P_{r_2,i_\varphi+s})
	+ \sum_{i_\varphi=1}^{s} \textrm{Cov}(N^P_{r_1,m_\varphi+i_\varphi-s},N^P_{r_2,i_\varphi}) \right) \nonumber \\
	\phantom{\sum_{i_\varphi=1}^{m_\varphi-s} =} \cdot \left( \sum_{i_\varphi=1}^{m_\varphi-s} \varepsilon_{r_1,i_\varphi} \varepsilon_{r_2,i_\varphi+s}
	+ \sum_{i_\varphi=1}^{s} \varepsilon_{r_1,m_\varphi+i_\varphi-s} \varepsilon_{r_2,i_\varphi} \right)
	\label{eq:twistsum}
\end{IEEEeqnarray}
Equation~(\ref{eq:twistsum}) uses the result in Eq.~(\ref{eq:DetectedCovVarSeg}) to relate the detected quantities to the primary quantities.
In the case here (where $s \geq 1$), the second piece of Eq.~(\ref{eq:DetectedCovVarSeg}) is always 0, because the terms never have the same $\varphi$.
Equation~(\ref{eq:twistsum}) can be inverted to allow one to compute the sum of invariant twisted covariances for primary particles from detected values.
\begin{IEEEeqnarray}{l}
	\sum_{i_\varphi=1}^{m_\varphi-s} \textrm{Cov}(N^P_{r_1,i_\varphi},N^P_{r_2,i_\varphi+s})
	+ \sum_{i_\varphi=1}^{s} \textrm{Cov}(N^P_{r_1,m_\varphi+i_\varphi-s},N^P_{r_2,i_\varphi}) = \nonumber \\
	\phantom{\sum_{i_\varphi=1}^{m_\varphi-s}} m_\varphi \cdot \frac{ \sum_{i_\varphi=1}^{m_\varphi-s} \textrm{Cov}(N^D_{r_1,i_\varphi},N^D_{r_2,i_\varphi+s})
	+ \sum_{i_\varphi=1}^{s} \textrm{Cov}(N^D_{r_1,m_\varphi+i_\varphi-s},N^D_{r_2,i_\varphi})}
	{\sum_{i_\varphi=1}^{m_\varphi-s} \varepsilon_{r_1,i_\varphi} \varepsilon_{r_2,i_\varphi+s}
	+ \sum_{i_\varphi=1}^{s} \varepsilon_{r_1,m_\varphi+i_\varphi-s} \varepsilon_{r_2,i_\varphi}}
	\label{eq:twistsuminverted}
\end{IEEEeqnarray}

The same analysis can be performed on the first term in Eq.~(\ref{eq:SumOfCovariancesTwists}), but now, when invoking Eq.~(\ref{eq:DetectedCovVarSeg}) the second piece must be kept as it may not vanish (when calculating a variance for example).
\begin{IEEEeqnarray}{rCl}
	\sum_{i_\varphi=1}^{m_\varphi} \textrm{Cov}(N^D_{r_1,i_\varphi},N^D_{r_2,i_\varphi}) &=&
	\sum_{i_\varphi=1}^{m_\varphi} \left( \textrm{Cov}(N^P_{r_1,i_\varphi},N^P_{r_2,i_\varphi}) \cdot \varepsilon_{r_1,i_\varphi} \varepsilon_{r_2,i_\varphi}
	\vphantom{+ \delta_{r_1 r_2} \cdot \langle N^P_{r_1,i_\varphi} \rangle \cdot \varepsilon_{r_1,i_\varphi} \left( 1 - \varepsilon_{r_1,i_\varphi} \right)} \right. \nonumber \\
	&& \hphantom{\sum_{i_\varphi=1}^{m_\varphi} \left( \vphantom{\textrm{Cov}(N^P_{r_1,i_\varphi},N^P_{r_2,i_\varphi}) \cdot \varepsilon_{r_1,i_\varphi} \varepsilon_{r_2,i_\varphi}
	+ \delta_{r_1 r_2} \cdot \langle N^P_{r_1,i_\varphi} \rangle \cdot \varepsilon_{r_1,i_\varphi} \left( 1 - \varepsilon_{r_1,i_\varphi} \right)} \right.}
	\left. \vphantom{\textrm{Cov}(N^P_{r_1,i_\varphi},N^P_{r_2,i_\varphi}) \cdot \varepsilon_{r_1,i_\varphi} \varepsilon_{r_2,i_\varphi}}
	+ \delta_{r_1 r_2} \cdot \langle N^P_{r_1,i_\varphi} \rangle \cdot \varepsilon_{r_1,i_\varphi} \left( 1 - \varepsilon_{r_1,i_\varphi} \right) \right) \nonumber \\
	&=& \frac{\sum_{i_\varphi=1}^{m_\varphi} \textrm{Cov}(N^P_{r_1,i_\varphi},N^P_{r_2,i_\varphi})}{m_\varphi} \cdot \sum_{i_\varphi=1}^{m_\varphi} \varepsilon_{r_1,i_\varphi} \varepsilon_{r_2,i_\varphi} \nonumber \\
	&& + \delta_{r_1 r_2} \cdot \frac{\sum_{i_\varphi=1}^{m_\varphi} \langle N^P_{r_1,i_\varphi} \rangle}{m_\varphi} \cdot \sum_{i_\varphi=1}^{m_\varphi} \varepsilon_{r_1,i_\varphi} \left( 1 - \varepsilon_{r_1,i_\varphi} \right)
	\label{eq:nontwistsum}
\end{IEEEeqnarray} 
Equation~(\ref{eq:nontwistsum}) can similarly be inverted to compute the sum of the non-twisted portion of Eq.~(\ref{eq:SumOfCovariancesTwists}) for primary particles:
\begin{IEEEeqnarray}{rCl}
	\sum_{i_\varphi=1}^{m_\varphi} \textrm{Cov}(N^P_{r_1,i_\varphi},N^P_{r_2,i_\varphi}) &=&
	m_\varphi \cdot \frac{\sum_{i_\varphi=1}^{m_\varphi} \textrm{Cov}(N^D_{r_1,i_\varphi},N^D_{r_2,i_\varphi})}{\sum_{i_\varphi=1}^{m_\varphi} \varepsilon_{r_1,i_\varphi} \varepsilon_{r_2,i_\varphi}} \nonumber \\
	&& - \delta_{r_1 r_2} \cdot m_\varphi \cdot \frac{\sum_{i_\varphi=1}^{m_\varphi} \varepsilon_{r_1,i_\varphi} \left( 1 - \varepsilon_{r_1,i_\varphi} \right)}{\sum_{i_\varphi=1}^{m_\varphi} \varepsilon^2_{r_1,i_\varphi}} \cdot \frac{\sum_{i_\varphi=1}^{m_\varphi} \langle N^D_{r_1,i_\varphi} \rangle}{\sum_{i_\varphi=1}^{m_\varphi} \varepsilon_{r_1,i_\varphi}}
	\label{eq:nontwistsuminverted}
\end{IEEEeqnarray} 
where Eq.~(\ref{eq:FinalMeanFormula}) was used to relate the mean number of primary particles to the mean number of detected particles.

The final expression for the covariance of primary particles is obtained by inserting Eqs.~(\ref{eq:twistsuminverted}) and (\ref{eq:nontwistsuminverted}) into Eq.~(\ref{eq:SumOfCovariancesTwists}):
\begin{IEEEeqnarray}{l}
	\textrm{Cov}(N^P_{r_1},N^P_{r_2}) = \nonumber \\
	\phantom{\textrm{Cov}} m_\varphi \cdot \frac{\sum_{i_\varphi=1}^{m_\varphi} \textrm{Cov}(N^D_{r_1,i_\varphi},N^D_{r_2,i_\varphi})}{\sum_{i_\varphi=1}^{m_\varphi} \varepsilon_{r_1,i_\varphi} \varepsilon_{r_2,i_\varphi}} \nonumber \\
	\phantom{\textrm{Cov}} + m_\varphi \cdot \sum_{s=1}^{m_\varphi-1} \left\{ \frac{ \sum_{i_\varphi=1}^{m_\varphi-s} \textrm{Cov}(N^D_{r_1,i_\varphi},N^D_{r_2,i_\varphi+s})
        + \sum_{i_\varphi=1}^{s} \textrm{Cov}(N^D_{r_1,m_\varphi+i_\varphi-s},N^D_{r_2,i_\varphi})}
        {\sum_{i_\varphi=1}^{m_\varphi-s} \varepsilon_{r_1,i_\varphi} \varepsilon_{r_2,i_\varphi+s}
        + \sum_{i_\varphi=1}^{s} \varepsilon_{r_1,m_\varphi+i_\varphi-s} \varepsilon_{r_2,i_\varphi}} \right\} \nonumber \\
	\phantom{\textrm{Cov}} - \delta_{r_1 r_2} \cdot m_\varphi \cdot \frac{\sum_{i_\varphi=1}^{m_\varphi} \varepsilon_{r_1,i_\varphi} \left( 1 - \varepsilon_{r_1,i_\varphi} \right)}{\sum_{i_\varphi=1}^{m_\varphi} \varepsilon^2_{r_1,i_\varphi}} \cdot \frac{\sum_{i_\varphi=1}^{m_\varphi} \langle N^D_{r_1,i_\varphi} \rangle}{\sum_{i_\varphi=1}^{m_\varphi} \varepsilon_{r_1,i_\varphi}}
	\label{eq:FinalCovFormula}
\end{IEEEeqnarray}
While the result using Eq.~(\ref{eq:FinalCovFormula}) must deviate from the result obtained from the distribution of primary particles (due to the imperfect detector response resulting in partial information loss), tests show a vast improvement over using Eqs.~(\ref{eq:CorrectedCov}) and (\ref{eq:CorrectedVar}).
Results using Eq.~(\ref{eq:FinalCovFormula}) often agree within statistical error with the results obtained from the primary distribution as will be shown in section~\ref{sec:verification}.

\begin{figure}
\begin{minipage}{0.32\textwidth}
\includegraphics[width=0.9\textwidth]{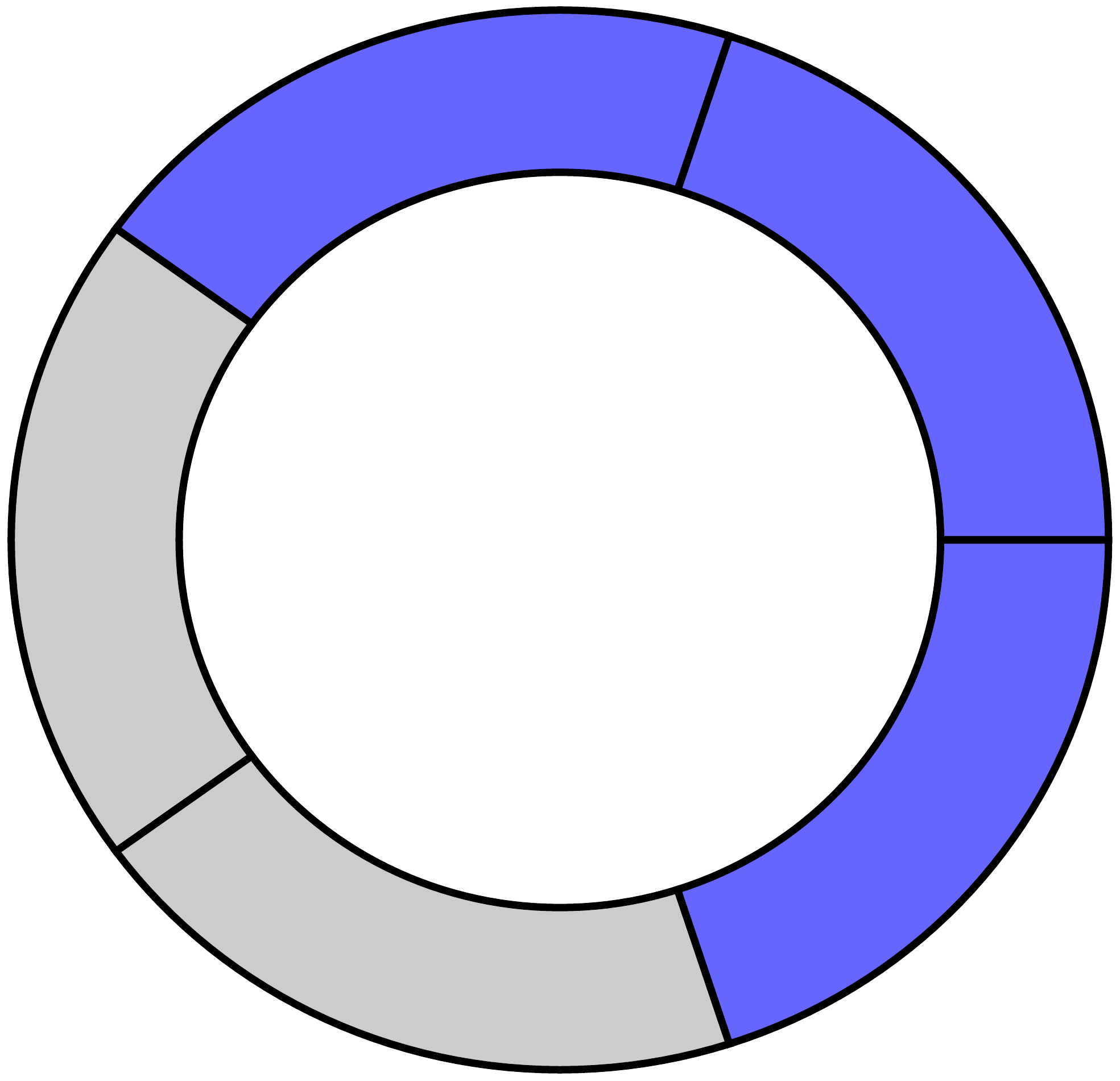}
\vspace*{-6mm}
\hphantom{\hspace{\textwidth}}
\begin{tabular}{|c|c|c|c|c|c|}
\hline
$s$ & 0 & 1 & 2 & 3 & 4\\
\hline
$N_{meas}$ & 3 & 2 & 1 & 1 & 2\\
\hline
scale factor & $\frac{5}{3}$ & $\frac{5}{2}$ & $\frac{5}{1}$ & $\frac{5}{1}$ & $\frac{5}{2}$\\
\hline
\end{tabular}
\end{minipage}
\begin{minipage}{0.32\textwidth}
\includegraphics[width=0.9\textwidth]{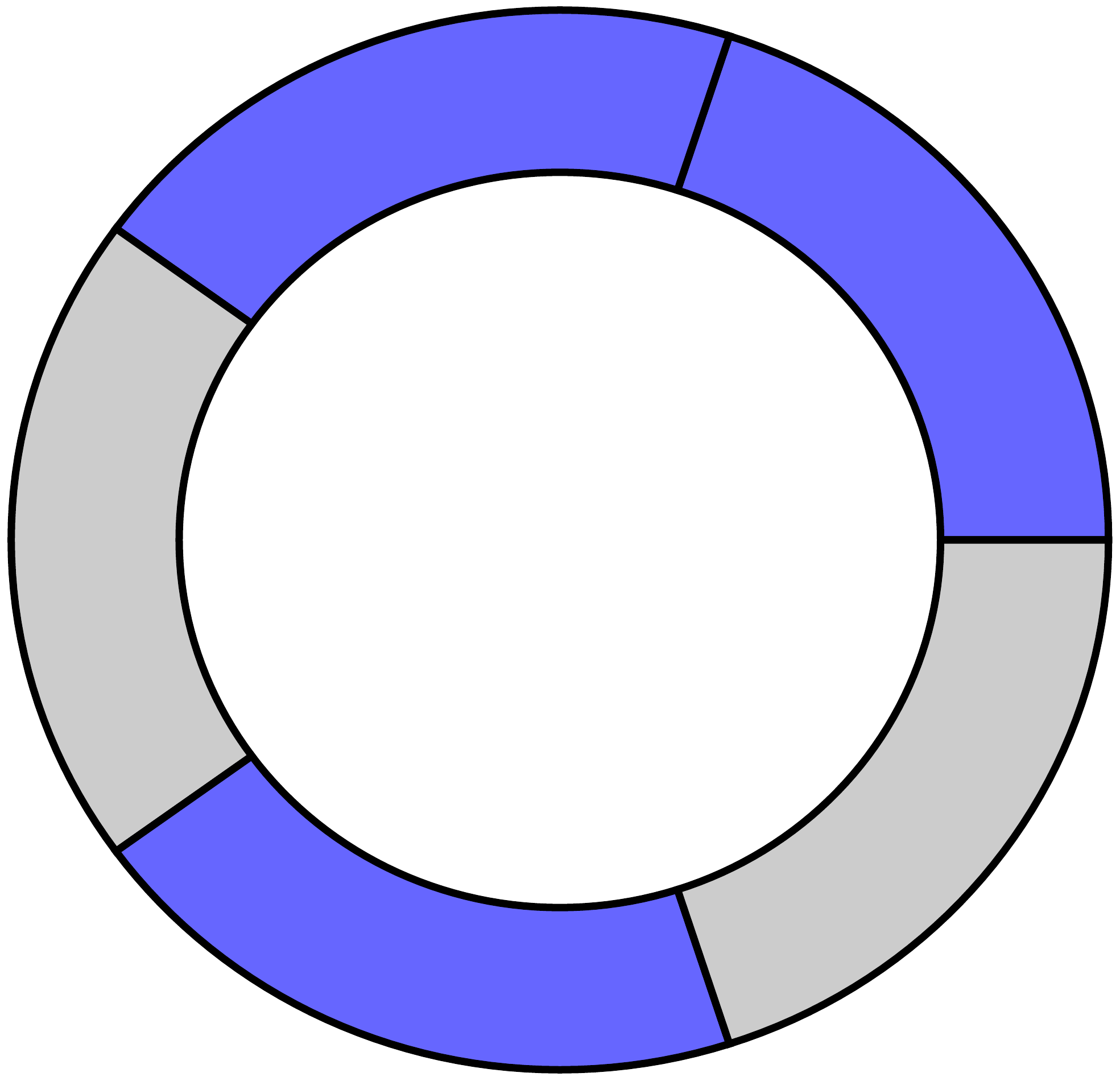}
\vspace*{-6mm}
\hphantom{\hspace{\textwidth}}
\begin{tabular}{|c|c|c|c|c|c|}
\hline
$s$ & 0 & 1 & 2 & 3 & 4\\
\hline
$N_{meas}$ & 3 & 1 & 2 & 2 & 1\\
\hline
scale factor & $\frac{5}{3}$ & $\frac{5}{1}$ & $\frac{5}{2}$ & $\frac{5}{2}$ & $\frac{5}{1}$\\
\hline
\end{tabular}
\end{minipage}
\begin{minipage}{0.32\textwidth}
\includegraphics[width=0.9\textwidth]{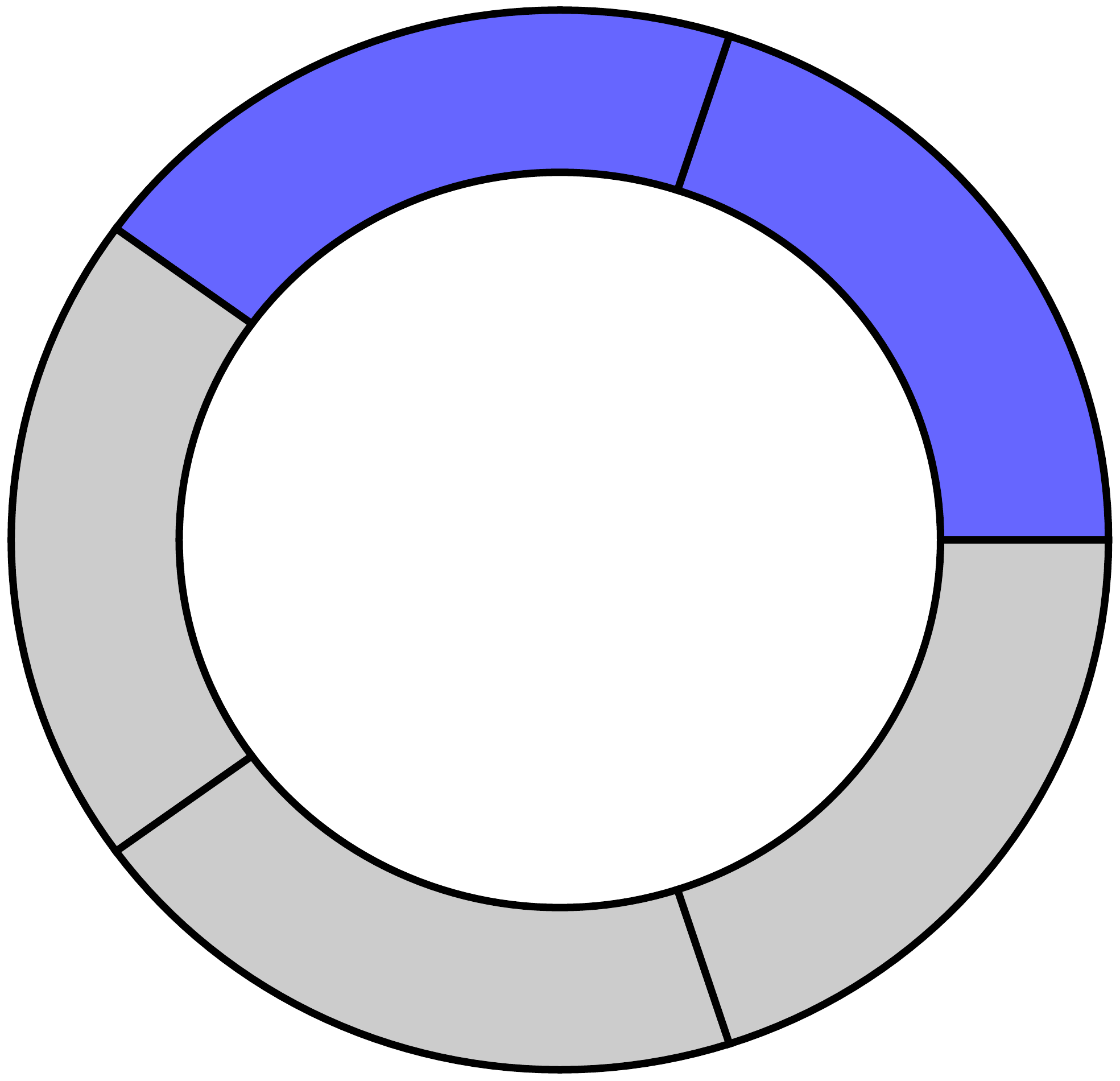}
\vspace*{-6mm}
\hphantom{\hspace{\textwidth}}
\begin{tabular}{|c|c|c|c|c|c|}
\hline
$s$ & 0 & 1 & 2 & 3 & 4\\
\hline
$N_{meas}$ & 2 & 1 & 0 & 0 & 1\\
\hline
scale factor & $\frac{5}{2}$ & $\frac{5}{1}$ & $\frac{5}{0}$ & $\frac{5}{0}$ & $\frac{5}{1}$\\
\hline
\end{tabular}
\end{minipage}
\caption{(Color online) The figures depict sets of dead regions (colored gray) for a detector that has 5 $\varphi$ segments.
If one assumes the same dead regions apply to both the forward and backward pseudorapidity regions (as an example), one can compute the number of times each invariant twist is measured ($N_{meas}$) for each configuration with a shift of $s$ segments and compute a scale factor that must be applied to give the correct contribution to the variance or covariance.
For the same total dead area one can still obtain different variances or covariances, because the configuration of dead regions affects the measurement through different scale factors applied to each invariant twist (shown in the left and middle panes).
The right pane shows that if one has less than 50\% acceptance, one will lack a measure of an invariant twist making the measurement impossible to calculate with this method.}
\label{fig:DeadConfig}
\end{figure}

If one considers the situation where each region has either full or zero acceptance, the denominators in the first two terms of Eq.~(\ref{eq:FinalCovFormula}) count the number of times the twisted (or non-twisted) quantities are measured.
The number of $\varphi$ segments divided by that number gives how much that quantity must be scaled up to give the appropriate contribution to covariance between the two region over $2\pi$ in azimuth. This is shown in Fig.~\ref{fig:DeadConfig}.
One should note that there is a limitation to this method that greater than 50\% of the acceptance must be present in each of the two regions being correlated.
If this requirement is not satisfied, one or more of the denominators summing over multiplications of efficiency factors in Eqs.~(\ref{eq:twistsuminverted}) and (\ref{eq:nontwistsuminverted}) will be 0.
This is a direct result of applying only rotational invariance to arrive at Eq.~(\ref{eq:FinalCovFormula}).
One may be able to lift this constraint by assuming that the invariant twists for primary particles shifted by $s$ segments are the same (on average) to the ones shifted by $m_\varphi-s$ segments.
Applying this symmetry was, however, not investigated further in this paper.

\subsection{Including an $\eta$ Dependent Efficiency}
\label{sec:etagrad}
If the efficiency additionally depends on $\eta$, the calculation of the correlation factor will also be affected.
In this case, the solution (if possible) is again to segment the detector (this time along $\eta$).
No redundancy necessarily exists in $\eta$ though, so one must be able to measure the variance of these sub-regions and the covariance of each sub-region to every other sub-region to accurately compute the correlation factor.
This means that each individual $\eta$ sub-region must have greater than 50\% acceptance.
The resulting equations are quite similar to those found in section~\ref{sec:eventshape}.

To specify the sub-region, a further subscript must be added to the primary and measured multiplicity to specify which $\eta$ and $\varphi$ segment is being referred to.
In extending Eq.~(\ref{eq:FinalMeanFormula}), the region is divided into $m_\eta$ $\eta$ segments.
This produces the following result:
\begin{IEEEeqnarray}{rCl}
	\langle N^P_r \rangle &=& m_\varphi \cdot \sum_{i_\eta=1}^{m_\eta} \frac{\sum_{i_\varphi=1}^{m_\varphi}\langle N^D_{r,i_\eta,i_\varphi}\rangle}{\sum_{i_\varphi=1}^{m_\varphi}\varepsilon_{r,i_\eta,i_\varphi}}
	\label{eq:FinalMeanFormulaEta}
\end{IEEEeqnarray}

For the covariance and the variances, region 1 and 2 will be segmented further into $m_{\eta,1}$ and $m_{\eta,2}$ $\eta$ segments, respectively.
This results in Eq.~(\ref{eq:DetectedCovVarSeg}) becoming
\begin{IEEEeqnarray}{rCl}
	\textrm{Cov}(N^D_{r_1,i_\eta,i_\varphi},N^D_{r_2,j_\eta,j_\varphi}) &=&
	\textrm{Cov}(N^P_{r_1,i_\eta,i_\varphi},N^P_{r_2,j_\eta,j_\varphi}) \cdot \varepsilon_{r_1,i_\eta,i_\varphi} \varepsilon_{r_2,j_\eta,j_\varphi} \nonumber\\
	&& + \delta_{r_1 r_2} \cdot \delta_{i_\eta j_\eta} \cdot \delta_{i_\varphi j_\varphi} \cdot \langle N^P_{r_1,i_\eta,i_\varphi} \rangle \cdot \varepsilon_{r_1,i_\eta,i_\varphi} \left( 1 - \varepsilon_{r_1,i_\eta,i_\varphi} \right)
	\label{eq:DetectedCovVarSegEta}
\end{IEEEeqnarray}
and Eq.~(\ref{eq:SumOfCovariancesTwists}) correspondingly becomes
\begin{IEEEeqnarray}{rCl}
	\textrm{Cov}(N^P_{r_1},N^P_{r_2}) &=& \sum_{i_\eta=1}^{m_{\eta,1}} \sum_{j_\eta=1}^{m_{\eta,2}} \sum_{i_\varphi=1}^{m_\varphi} \sum_{j_\varphi=1}^{m_\varphi} \textrm{Cov}(N^P_{r_1,i_\eta,i_\varphi},N^P_{r_2,j_\eta,j_\varphi}) \nonumber \\
	&=& \sum_{i_\eta=1}^{m_{\eta,1}} \sum_{j_\eta=1}^{m_{\eta,2}} \sum_{i_\varphi=1}^{m_\varphi} \textrm{Cov}(N^P_{r_1,i_\eta,i_\varphi},N^P_{r_2,j_\eta,i_\varphi}) \nonumber \\
	&& + \sum_{i_\eta=1}^{m_{\eta,1}} \sum_{j_\eta=1}^{m_{\eta,2}} \sum_{s=1}^{m_\varphi-1} \left\{
	\sum_{i_\varphi=1}^{m_\varphi-s} \textrm{Cov}(N^P_{r_1,i_\eta,i_\varphi},N^P_{r_2,j_\eta,i_\varphi+s})
	\vphantom{+ \sum_{i_\varphi=1}^{s} \textrm{Cov}(N^P_{r_1,i_\eta,m_\varphi+i_\varphi-s},N^P_{r_2,j_\eta,i_\varphi})}
	\right. \nonumber \\
	&& \hphantom{+ \sum_{i_\eta=1}^{m_{\eta,1}} \sum_{j_\eta=1}^{m_{\eta,2}} \sum_{s=1}^{m_\varphi-1} \left\{
	\vphantom{\sum_{i_\varphi=1}^{m_\varphi-s} \textrm{Cov}(N^P_{r_1,i_\eta,i_\varphi},N^P_{r_2,j_\eta,i_\varphi+s})
	+ \sum_{i_\varphi=1}^{s} \textrm{Cov}(N^P_{r_1,i_\eta,m_\varphi+i_\varphi-s},N^P_{r_2,j_\eta,i_\varphi})}
	\right.}
	\left.
	\vphantom{\sum_{i_\varphi=1}^{m_\varphi-s} \textrm{Cov}(N^P_{r_1,i_\eta,i_\varphi},N^P_{r_2,j_\eta,i_\varphi+s})}
	+ \sum_{i_\varphi=1}^{s} \textrm{Cov}(N^P_{r_1,i_\eta,m_\varphi+i_\varphi-s},N^P_{r_2,j_\eta,i_\varphi})
	\right\}
	\label{eq:SumOfCovariancesTwistsEta}
\end{IEEEeqnarray}

Equations~(\ref{eq:twistsuminverted}) and (\ref{eq:nontwistsuminverted}) apply to each $\eta$ segment pair and, therefore, the final formula incorporating both a $\varphi$ and $\eta$ efficiency gradient is the following:
\begin{IEEEeqnarray}{l}
	\textrm{Cov}(N^P_{r_1},N^P_{r_2}) = \nonumber \\
	\phantom{\textrm{Cov}} m_\varphi \cdot \sum_{i_\eta=1}^{m_{\eta,1}} \sum_{j_\eta=1}^{m_{\eta,2}} \frac{\sum\limits_{i_\varphi=1}^{m_\varphi} \textrm{Cov}(N^D_{r_1,i_\eta,i_\varphi},N^D_{r_2,j_\eta,i_\varphi})}{\sum\limits_{i_\varphi=1}^{m_\varphi} \varepsilon_{r_1,i_\eta,i_\varphi} \varepsilon_{r_2,j_\eta,i_\varphi}} \nonumber \\
	\phantom{\textrm{Cov}} + m_\varphi \cdot \sum_{i_\eta=1}^{m_{\eta,1}} \sum_{j_\eta=1}^{m_{\eta,2}} \sum_{s=1}^{m_\varphi-1} \left\{ \frac{ \sum\limits_{i_\varphi=1}^{m_\varphi-s} \textrm{Cov}(N^D_{r_1,i_\eta,i_\varphi},N^D_{r_2,j_\eta,i_\varphi+s})
        + \sum\limits_{i_\varphi=1}^{s} \textrm{Cov}(N^D_{r_1,i_\eta,m_\varphi+i_\varphi-s},N^D_{r_2,j_\eta,i_\varphi})}
        {\sum\limits_{i_\varphi=1}^{m_\varphi-s} \varepsilon_{r_1,i_\eta,i_\varphi} \varepsilon_{r_2,j_\eta,i_\varphi+s}
        + \sum\limits_{i_\varphi=1}^{s} \varepsilon_{r_1,i_\eta,m_\varphi+i_\varphi-s} \varepsilon_{r_2,j_\eta,i_\varphi}} \right\} \nonumber \\
	\phantom{\textrm{Cov}} - \delta_{r_1 r_2} \cdot m_\varphi \cdot \sum_{i_\eta=1}^{m_{\eta,1}} \frac{\sum\limits_{i_\varphi=1}^{m_\varphi} \varepsilon_{r_1,i_\eta,i_\varphi} \left( 1 - \varepsilon_{r_1,i_\eta,i_\varphi} \right)}{\sum\limits_{i_\varphi=1}^{m_\varphi} \varepsilon^2_{r_1,i_\eta,i_\varphi}} \cdot \frac{\sum\limits_{i_\varphi=1}^{m_\varphi} \langle N^D_{r_1,i_\eta,i_\varphi} \rangle}{\sum\limits_{i_\varphi=1}^{m_\varphi} \varepsilon_{r_1,i_\eta,i_\varphi}}
	\label{eq:FinalCovFormulaEta}
\end{IEEEeqnarray}
Equation~(\ref{eq:FinalCovFormulaEta}) can also be used when the $\eta$ bin width for the final desired measurement is larger than the $\eta$ segmentation of the detector.
If the detector additionally has inactive channels that do not cover the full desired $\eta$ bin width, one can account for this using Eq.~(\ref{eq:FinalCovFormulaEta}) to further reduce measurement bias.

\section{Verification}
\label{sec:verification}
\begin{figure}[ht]
\begin{center}
\includegraphics[width=0.32\textwidth]{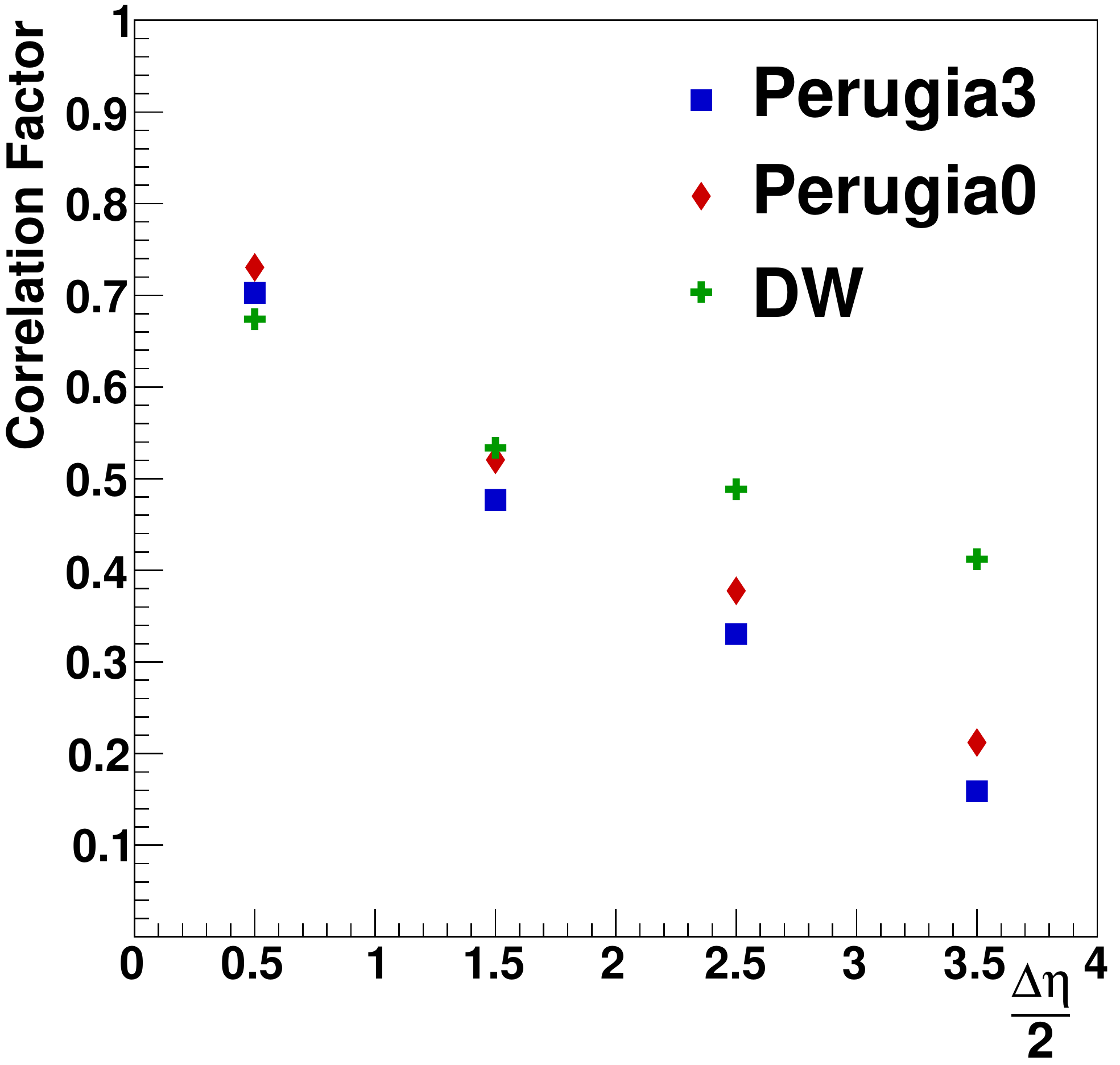}
\includegraphics[width=0.32\textwidth]{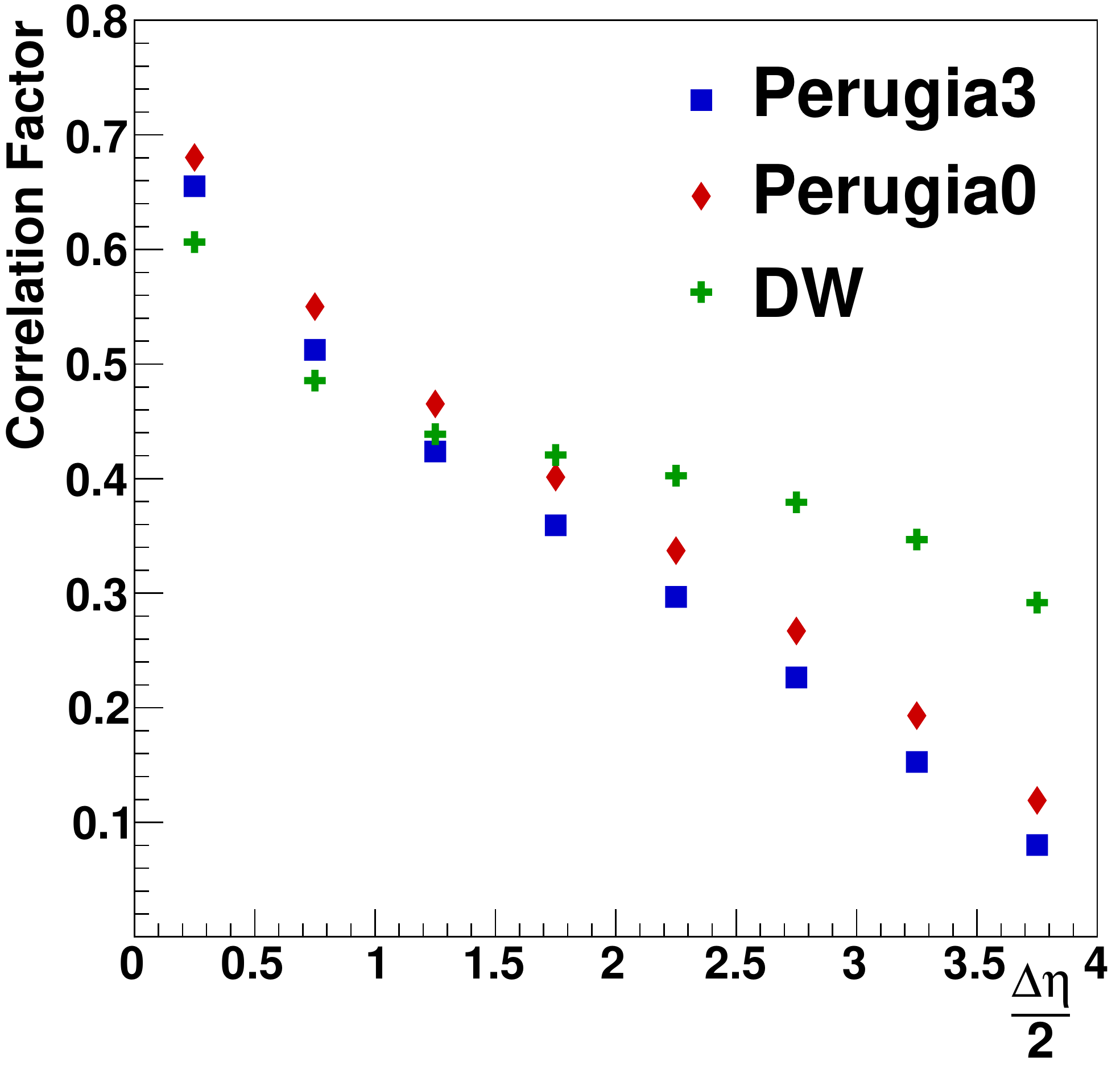}
\includegraphics[width=0.32\textwidth]{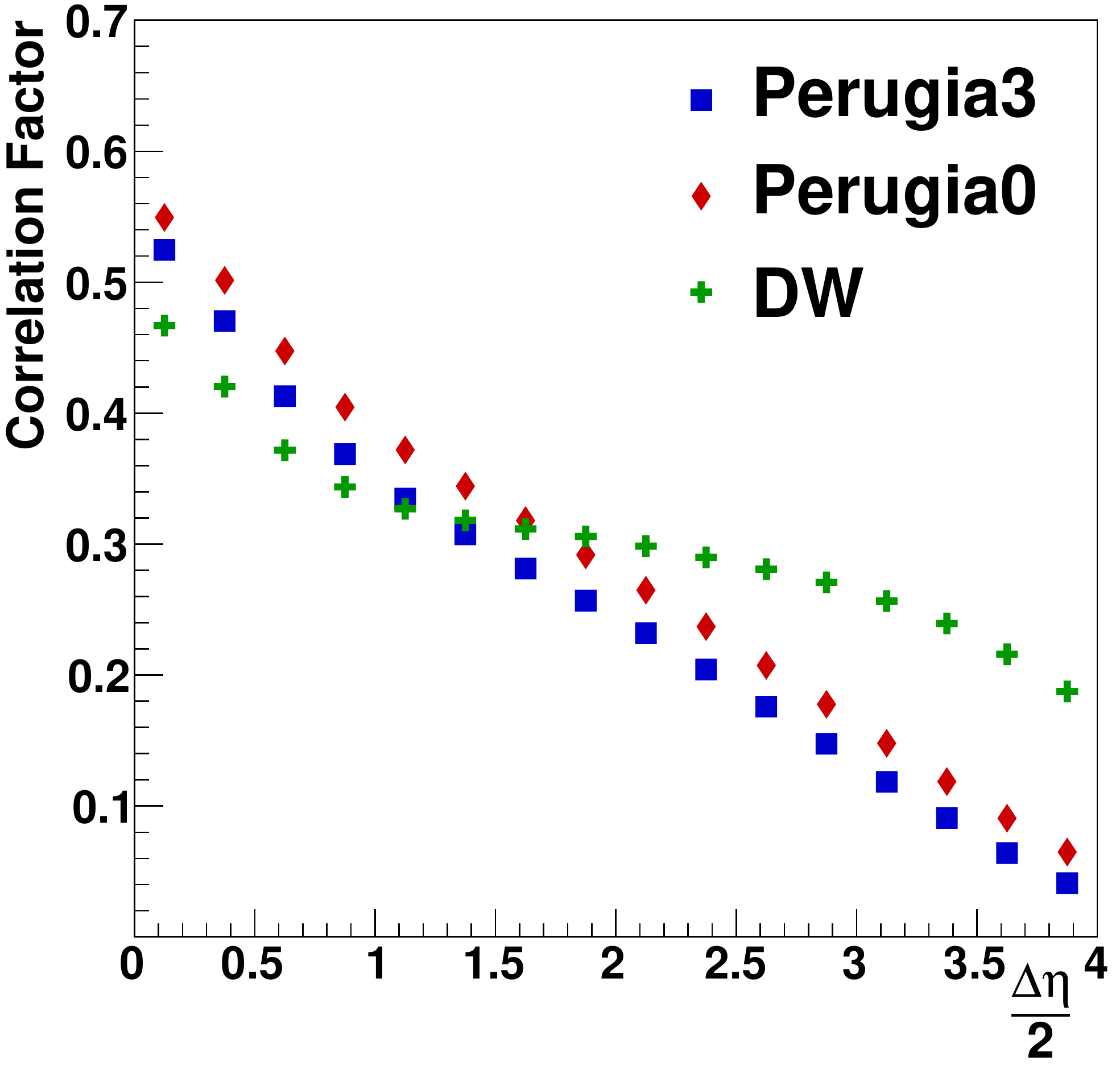}
\caption{(Color online) Correlation factors ($b$) obtained using the particle multiplicities computed at $\eta = \frac{\Delta\eta}{2}$ and $\eta = -\frac{\Delta\eta}{2}$ from the Pythia6 event-generator.
Three different tunes were selected: Perugia3, Perugia0, and DW.
Three different bin widths, $\Delta_{\textrm{bin}}$, in $\eta$ were used. Left: $\Delta_{\textrm{bin}}=1$. Middle: $\Delta_{\textrm{bin}}=0.5$. Right: $\Delta_{\textrm{bin}}=0.25$.}
\label{fig:PrimComp}
\end{center}
\end{figure}

The validity of the developed method is verified through studies using simulations of proton-proton collisions.
The event generator used here is Pythia 6.4~\cite{Sjostrand:2006za}.
It has been chosen, because many pre-configured tunes exist which predict substantially different quantities for different observables and specifically, in this case, for forward-backward correlations.
The properties of the tunes can be found in~\cite{Skands:2010ak}.
The tunes used for this study are Perugia3, Perugia0, and DW.
The DW tune produces quite different correlation factors when compared to the other two tunes (see Fig.~\ref{fig:PrimComp}) due to the significantly different relative contributions of initial state radiation compared to multiple parton interactions.
One should note that the bin width in $\eta$, $\Delta_{\textrm{bin}}$, affects the value of $b$ with $b \rightarrow 0$ as $\Delta_{\textrm{bin}} \rightarrow 0$.

\begin{figure}[ht]
\begin{center}
\includegraphics[width=0.32\textwidth]{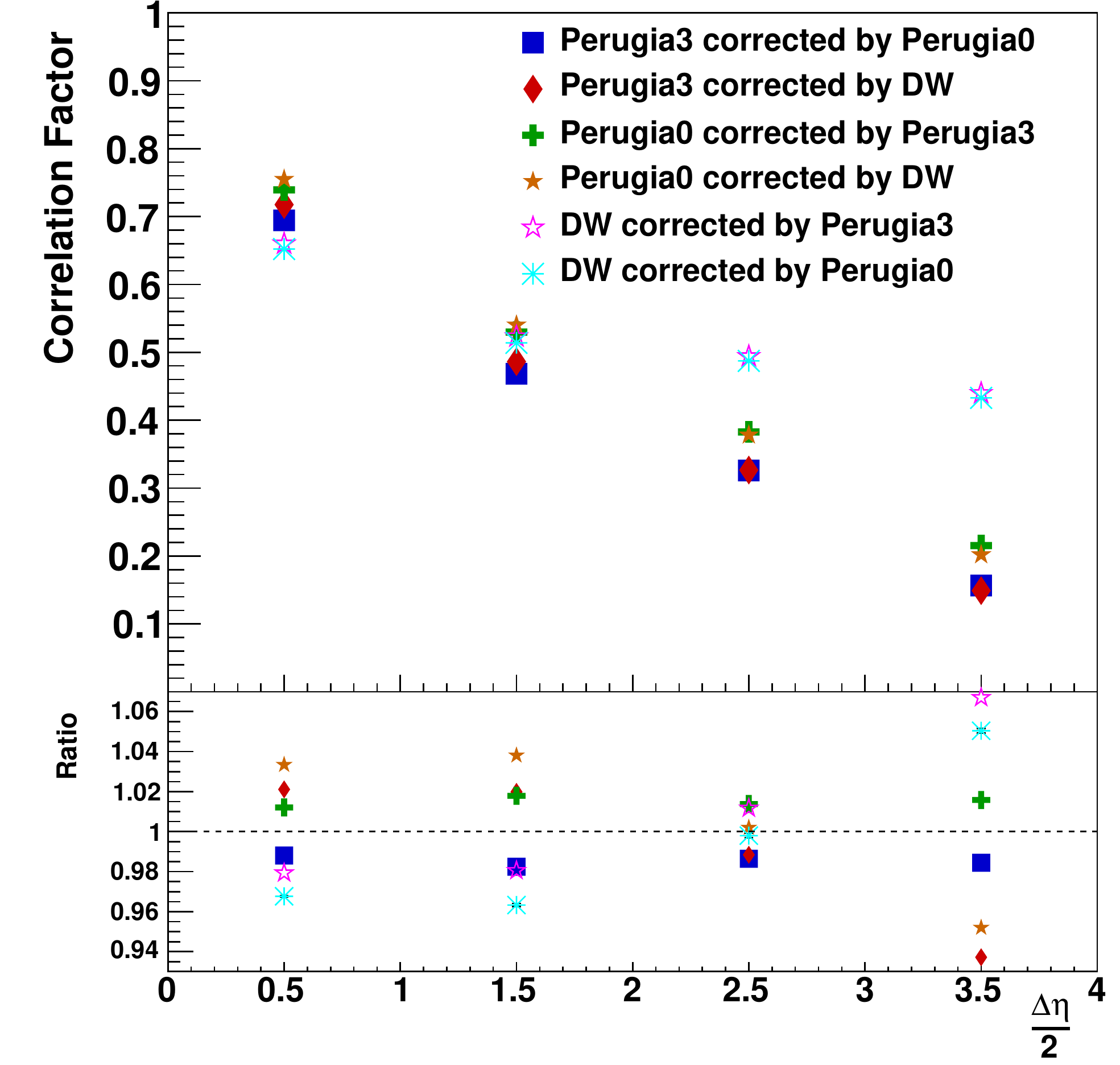}
\includegraphics[width=0.32\textwidth]{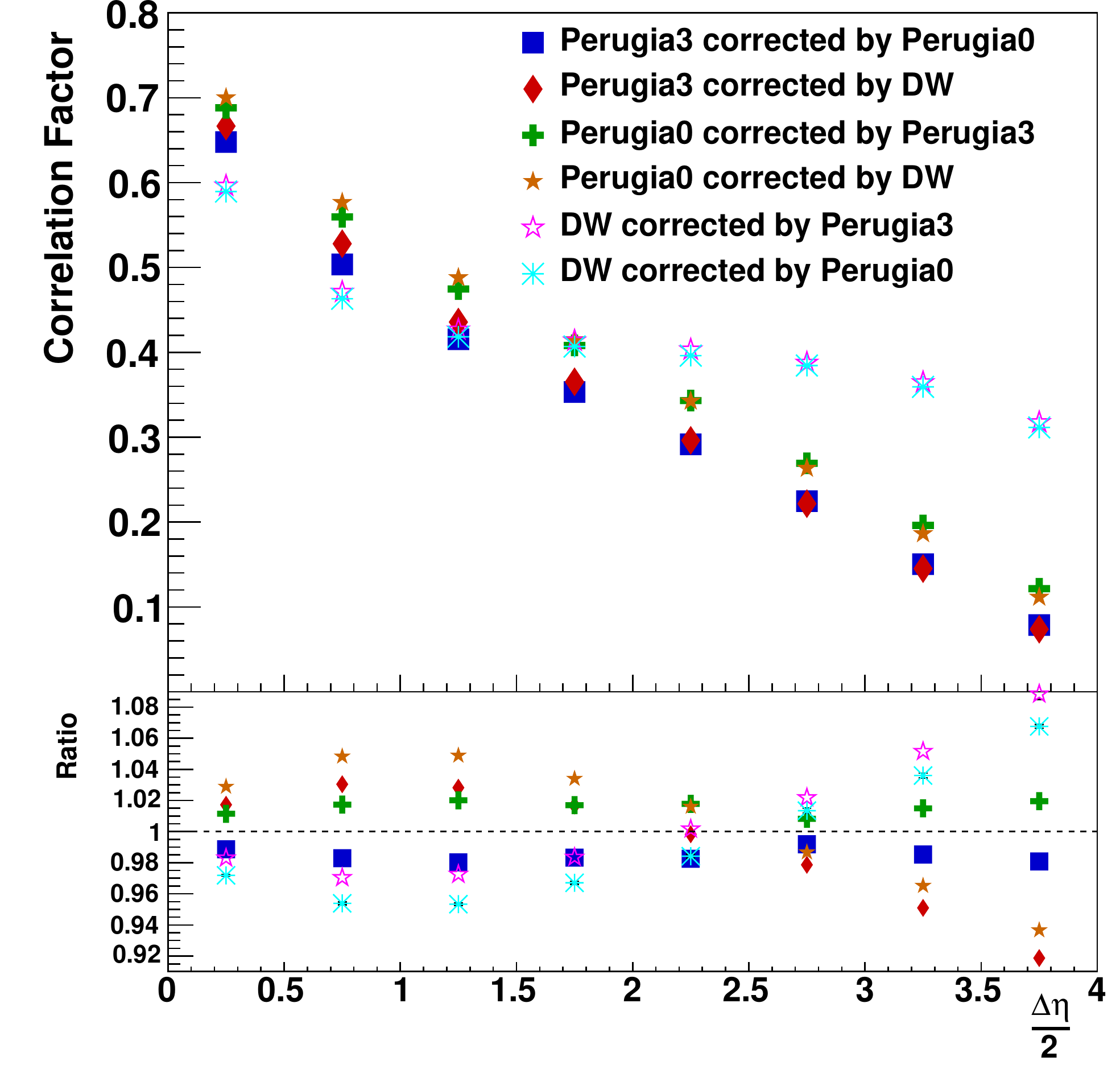}
\includegraphics[width=0.32\textwidth]{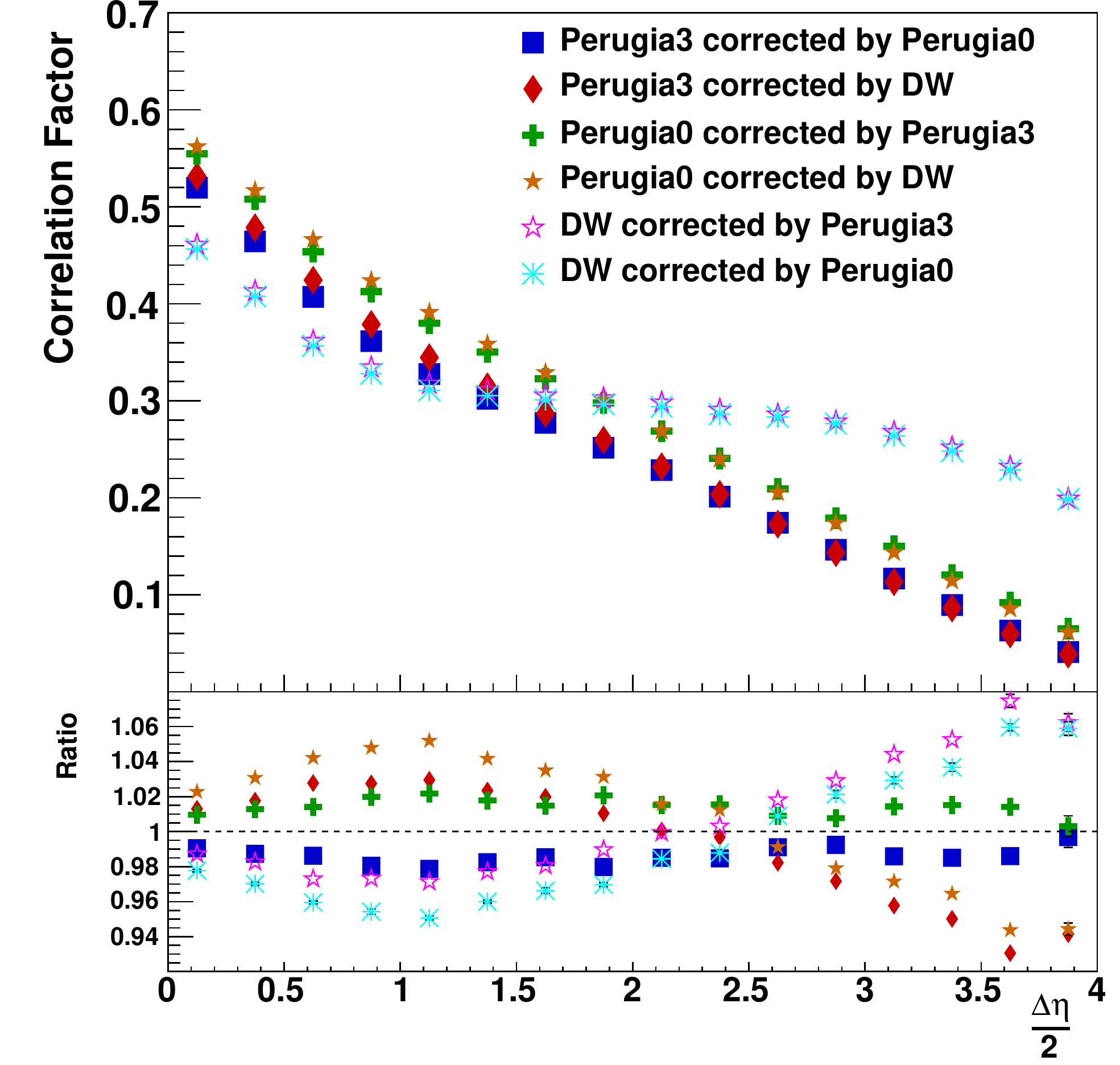}
\caption{(Color online) Correlation factors ($b$) obtained when only 60\% of the total acceptance exists in $\varphi$ equivalently over all $\eta$ correcting the found correlation factor from one Pythia6 tune by the ratio of the primary to found correlation factor found with another tune.
Three different bin widths, $\Delta_{\textrm{bin}}$, in $\eta$ were shown. Left: $\Delta_{\textrm{bin}}=1$. Middle: $\Delta_{\textrm{bin}}=0.5$. Right: $\Delta_{\textrm{bin}}=0.25$.
The bottom part of each figure shows the ratio of the corrected correlation factor to the actual primary correlation factor for the tune that the data was generated from.}
\label{fig:SimCorr}
\end{center}
\end{figure}

The most common method of accounting for a detector effect is to use simulated data to evaluate a quantity both with and without detector effects included.
The ratio is then used as a correction to the actual measured value which manifestly includes detector effects.
The validity of that method must, however, be assessed to establish if any residual dependence on the parameters of the generator exists.
As an example, the primary particles that were used to produce Fig.~\ref{fig:PrimComp} were subjected to a uniform contiguous acceptance hole of 40\% in $\varphi$ for all $\eta$ bins.
The detected correlation factors found with each tune were then corrected using the ratio of the true to detected factors found with the other tunes.
The result is shown in Fig.~\ref{fig:SimCorr}.
Deviations from the original primary correlation factors of up to 8\% are found in this case.
The deviations clearly show residual generator dependencies and biases.
Furthermore, real data could disagree even further with the tune chosen to correct with and could, therefore, produce even bigger biases.

In the following examples the simulation independent method developed in the previous section is used to evaluate the correlation factors.
Unless otherwise specified, the primary Pythia tune used in the examples is Perugia3.
Additionally, for all plots shown in the rest of this paper, $\Delta_{\textrm{bin}}=0.5$ is the bin width in $\eta$ unless the bin width is explicitly stated.

\subsection{Reduced Acceptance}
\label{sec:reducedacc}
The initial study involves the reduction of the acceptance of each $\eta$ bin.
Two examples are studied: a simple case, where inactive regions have identical $\varphi$ locations in all $\eta$ bins, and a realistic case, where inactive regions have been placed randomly into each $\eta$ bin.
In both cases, geometrical areas are chosen to be inactive with respect to particle detection, meaning that any particle with a momentum vector pointing toward an inactive region is excluded from the detected quantities. 

\subsubsection{The Simple Case}
\label{sec:simplecase}
\begin{figure}[ht]
\begin{center}
\includegraphics[width=0.4\textwidth]{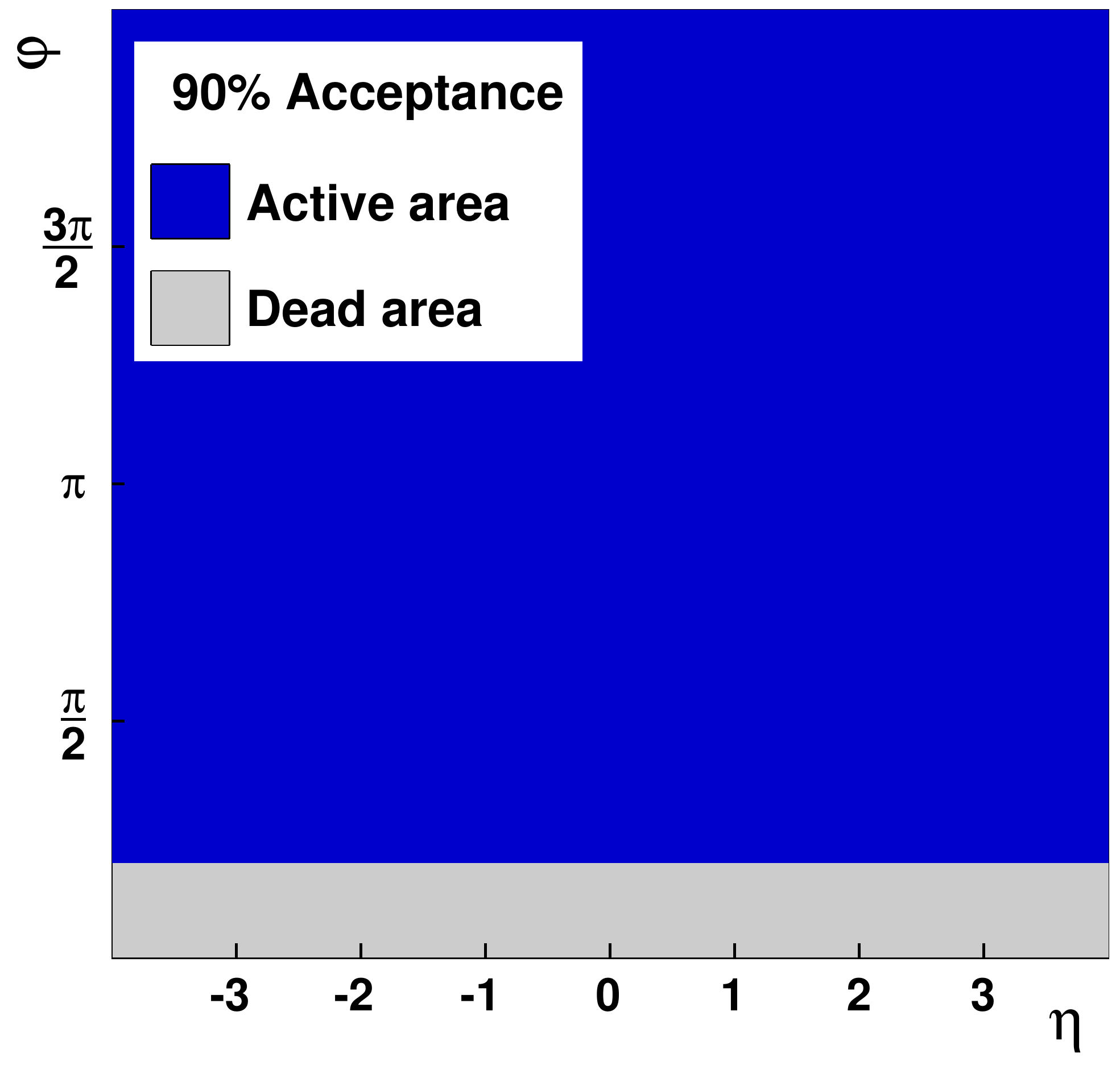}
\hspace{4ex}
\includegraphics[width=0.4\textwidth]{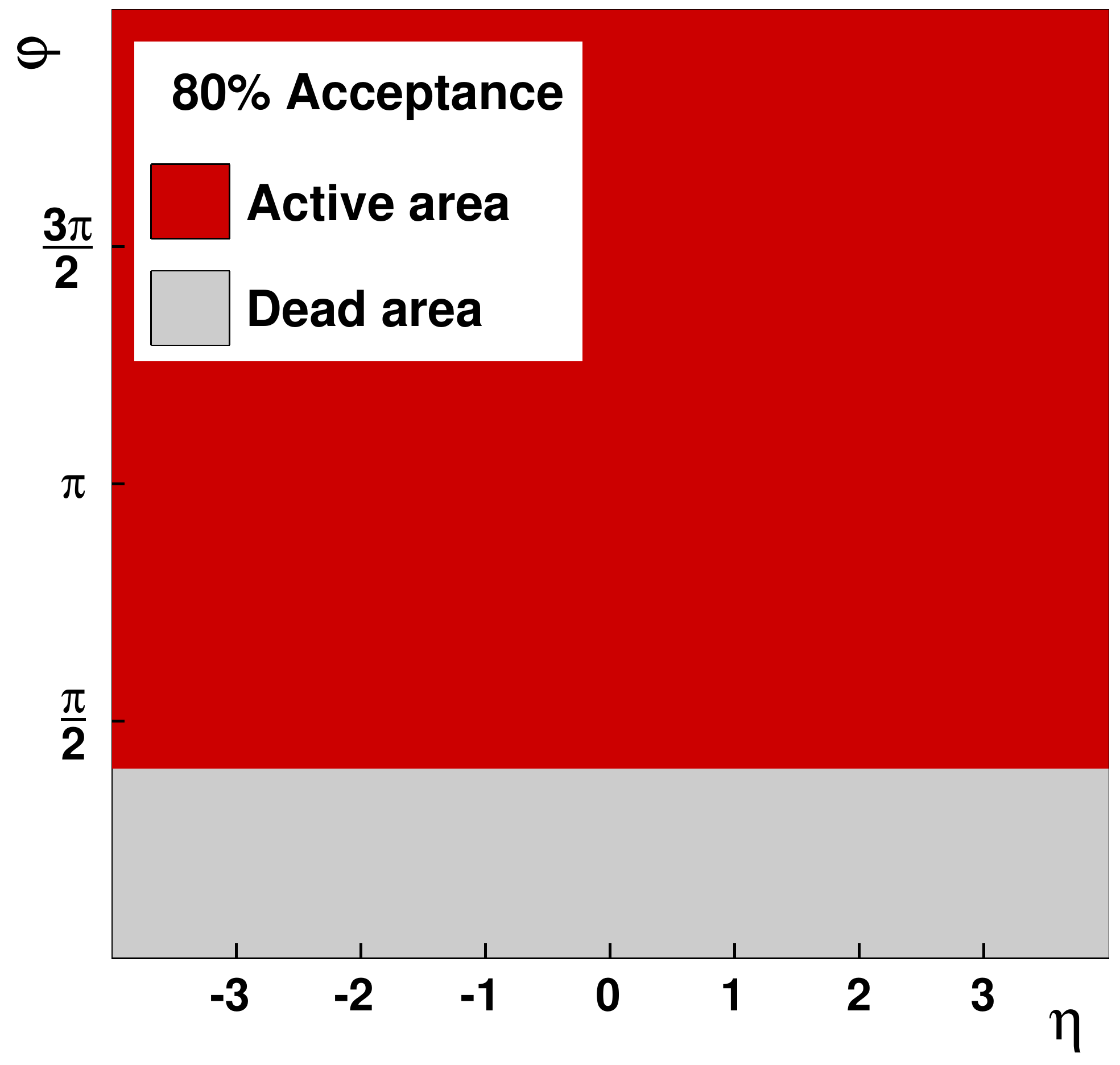}
\includegraphics[width=0.4\textwidth]{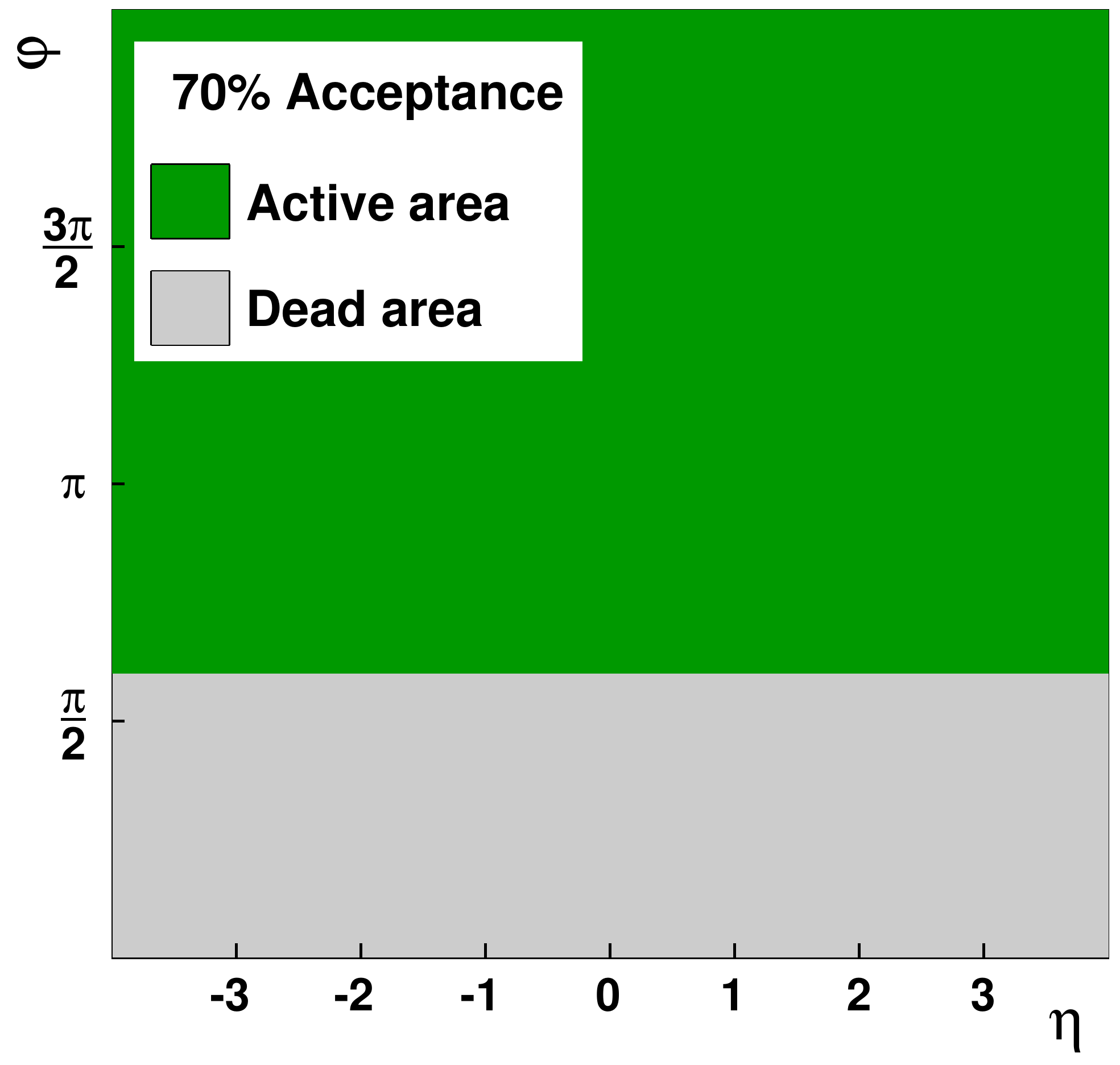}
\hspace{4ex}
\includegraphics[width=0.4\textwidth]{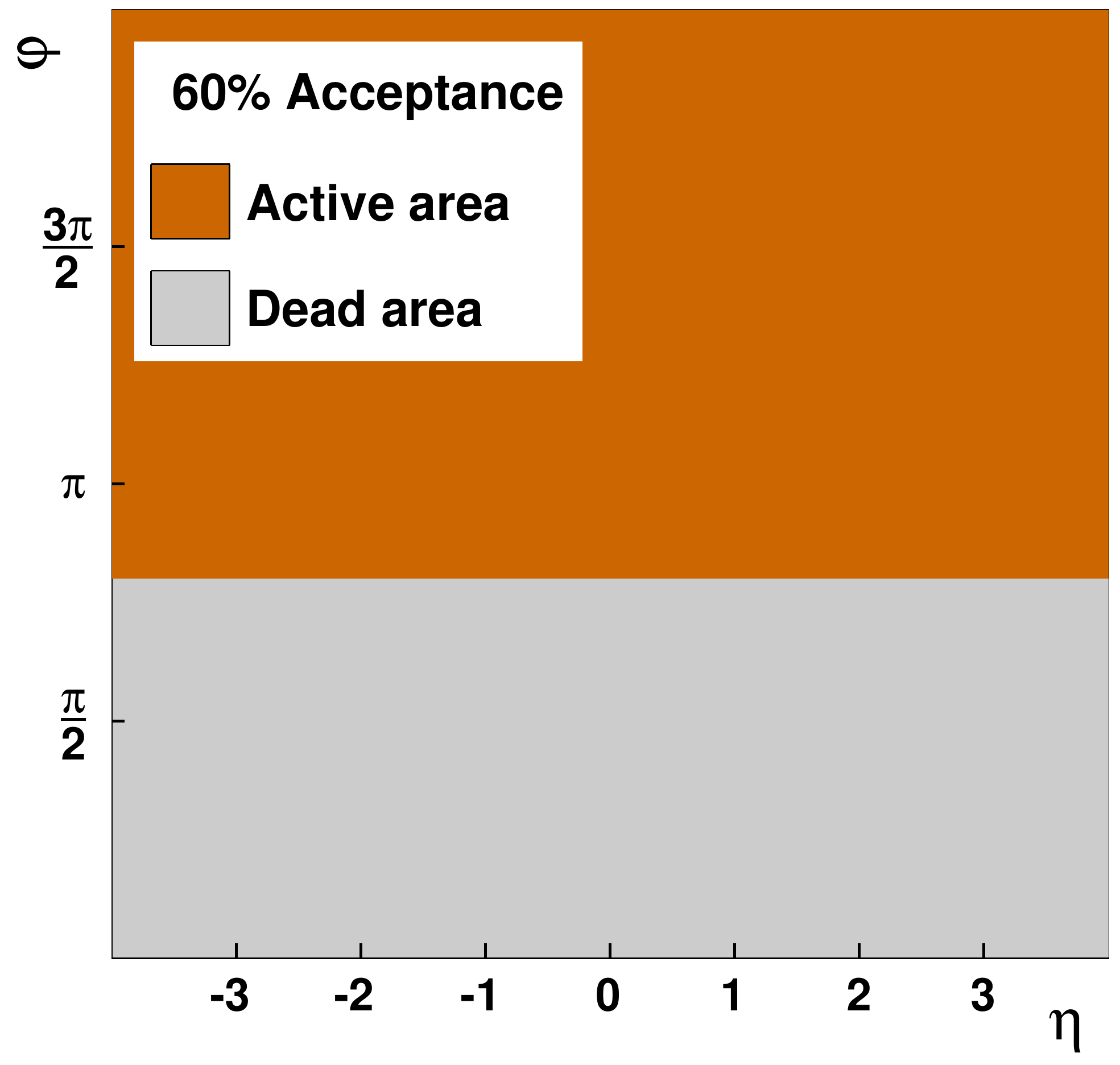}
\caption{(Color online) The four simple examples of the acceptance maps of the $\eta$ bins.
The four panes show the inactive regions with acceptances of 90\%, 80\%, 70\%, and 60\%.}
\label{fig:AccMapSym}
\end{center}
\end{figure}

Four simple examples are investigated in this section.
The inactive areas are chosen such that they begin at $\varphi=0$ and extend to $n\cdot\frac{2\pi}{10}$ where $n=1,2,3,\textrm{ and } 4$.
This results in geometric acceptances for each $\eta$ bin of 90\%, 80\%, 70\%, and 60\%.
The acceptance maps are shown in Fig.~\ref{fig:AccMapSym}.

\begin{figure}[ht]
\begin{center}
\includegraphics[width=0.49\textwidth]{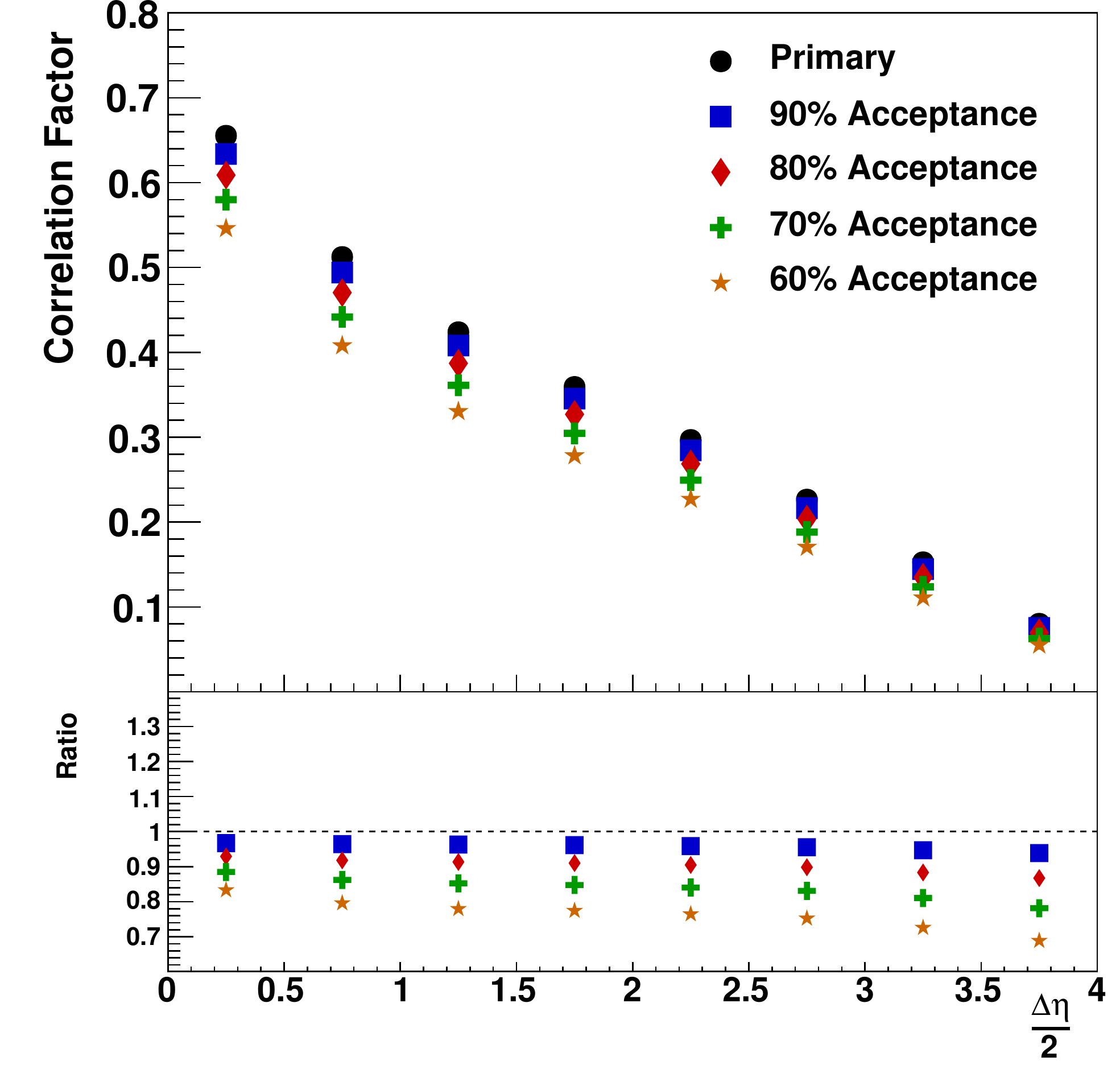}
\includegraphics[width=0.49\textwidth]{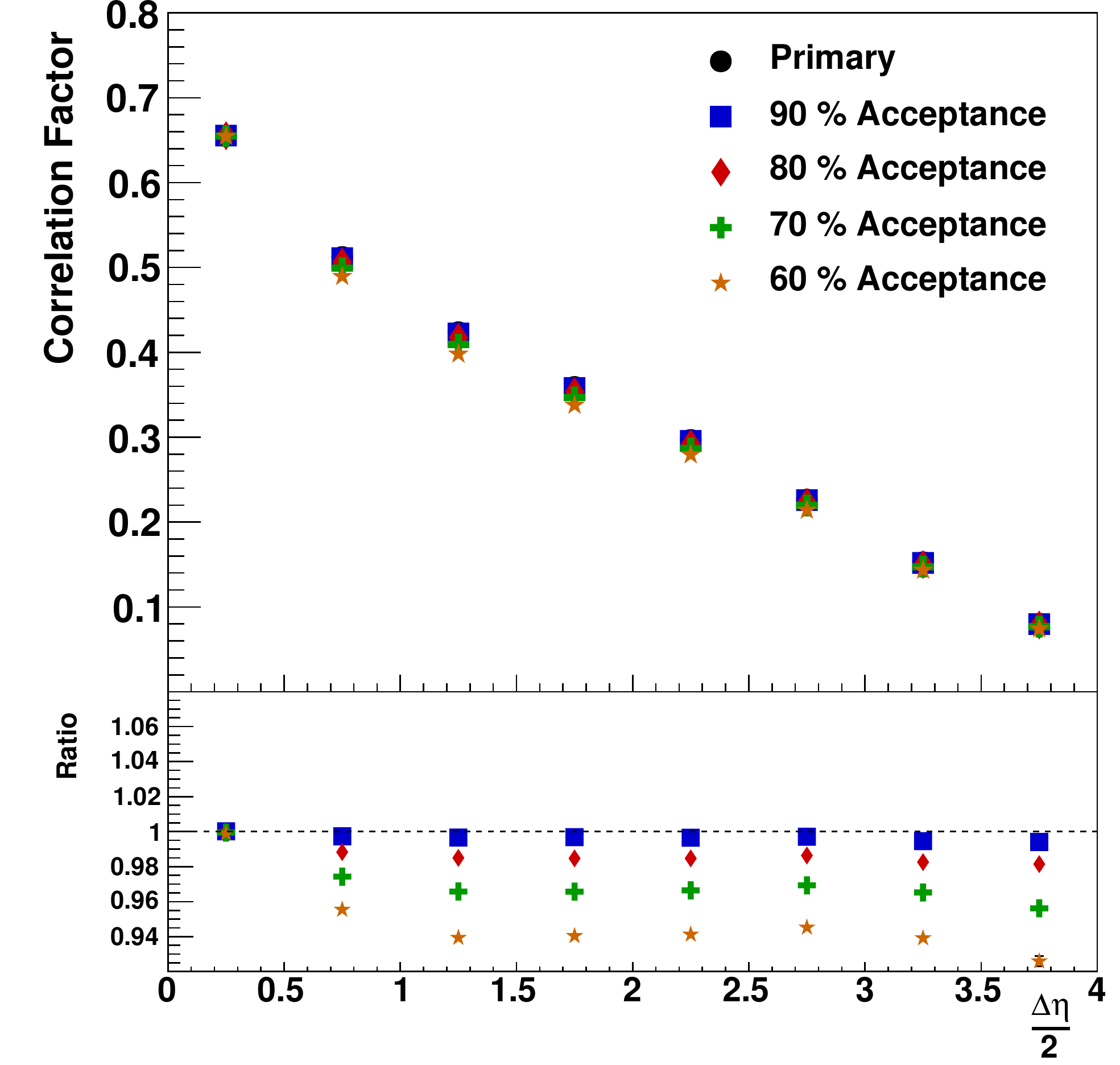}
\caption{(Color online) Left: Attenuation of measured correlation factors as a result of decreased $\varphi$ acceptance in the $\eta$ bins (see Fig.~\ref{fig:AccMapSym}).
Right: The obtained correlation factors computed using Eq.~(\ref{eq:BasicAccCorrection}), which does not exploit $\varphi$ segmentation.
Both: The bottom parts of the figures show the ratio of the obtained correlation factors to the primary correlation factors.
The simple assumption of uniform detection probability reduces the discrepancy from the primary correlation factor by more than a factor of 3, but still leaves substantial discrepancies between the obtained correlation factors and those from the generator output.}
\label{fig:VarAccP3Simp}
\end{center}
\end{figure}

Regardless of the cause, undetected particles will result in a loss of information and will affect the measured correlation factor.
We would intuitively expect that the correlation factors are attenuated when the efficiency of a bin is less than 1.
Equation~(\ref{eq:BasicAccCorrection}) demonstrates this.
This effect is illustrated in left pane of Fig.~\ref{fig:VarAccP3Simp}.
The correlation factor at the event-generator level is black while other colors are used for each case of reduced acceptance.
The graph shows that more attenuation exists when the size of the inactive areas is increased. 

\begin{figure}[ht]
\begin{center}
\includegraphics[width=0.49\textwidth]{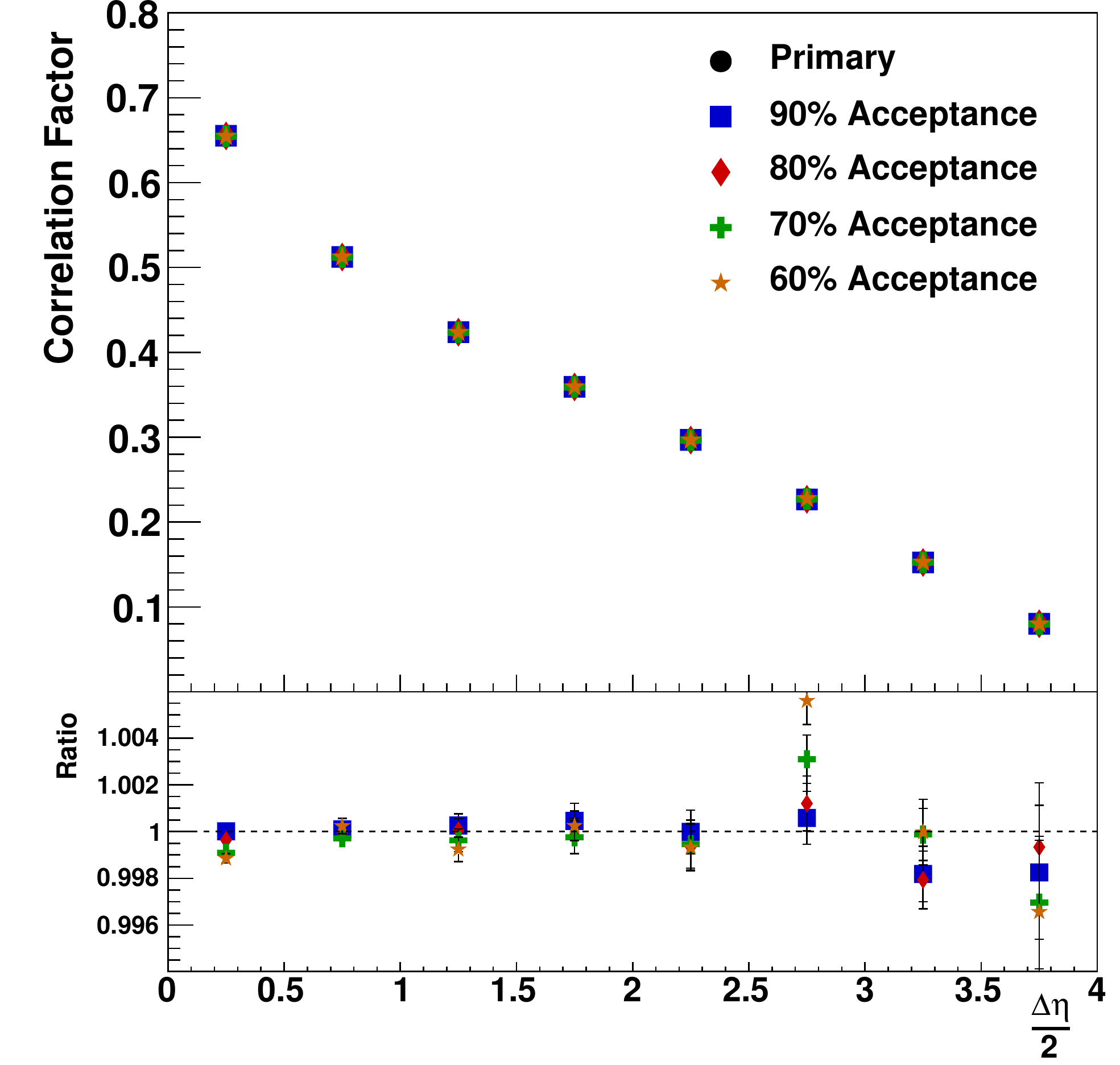}
\includegraphics[width=0.49\textwidth]{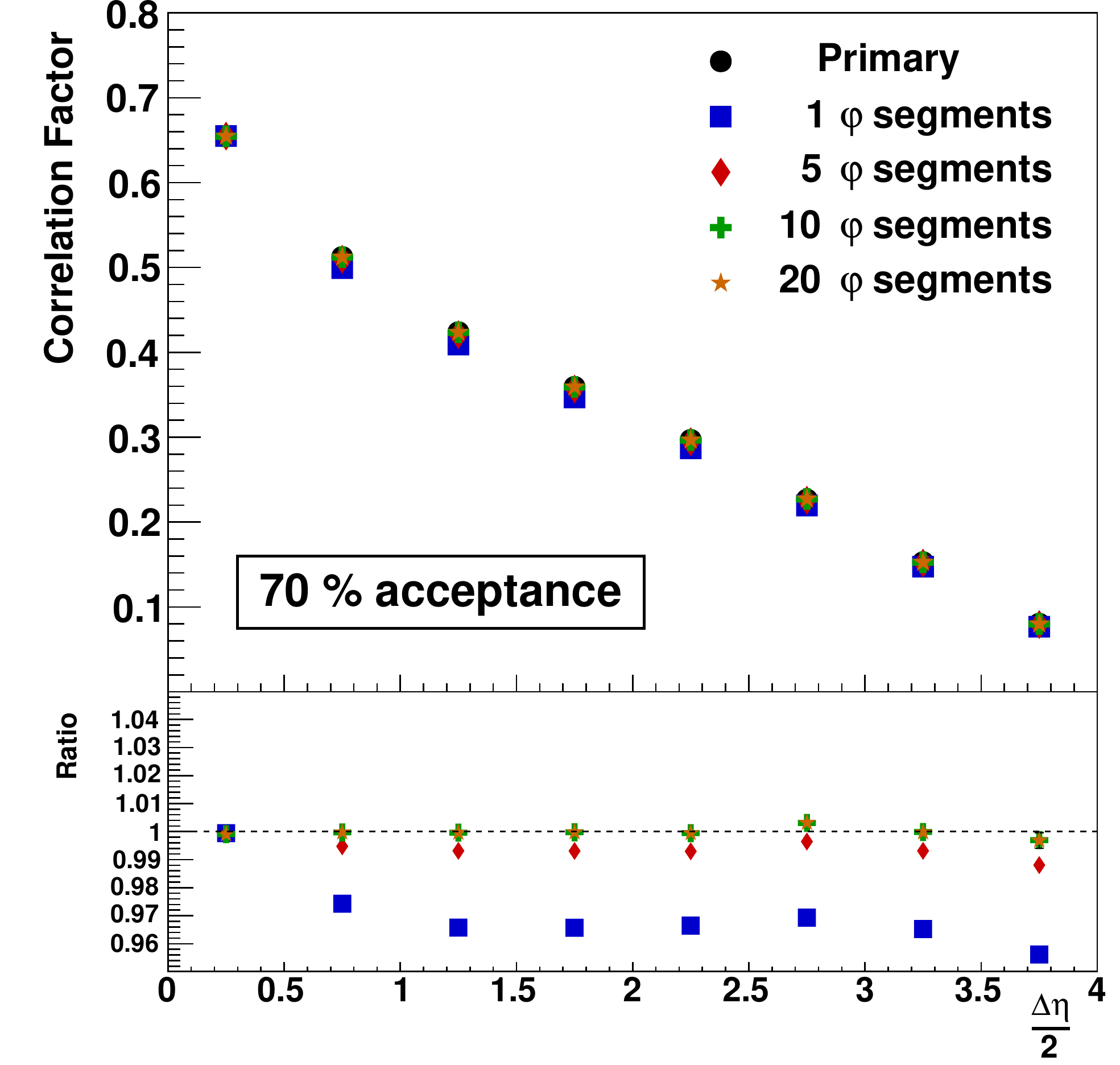}
\caption{(Color online) Left: Correlation factors computed using 10 $\varphi$ segments to account for the azimuthal event shape.
Right: Correlation factors obtained when 70\% acceptance and a varied number of $\varphi$ segments in each $\eta$ bin exists.
Note that there is no improvement when increasing from 10 to 20 $\varphi$ segments, because both segmentations produce the same result analytically for the acceptance map applied here.
Both: The bottom parts of the figures show the ratio of the obtained correlation factors to the primary correlation factors.
Note that the scale in the bottom part of the left figure has a much smaller range than that from the previous plots with no $\varphi$ segmentation.}
\label{fig:VarAccCorrP3Simp}
\end{center}
\end{figure}

To illustrate the necessity of segmentation, the results are first computed without any segmentation (using Eq.~(\ref{eq:BasicAccCorrection})).
The results are shown in the right pane of Fig.~\ref{fig:VarAccP3Simp}.
Although the computed correlation factors are now less than 10\% from the primary correlation factors, the discrepancy is still sizable.
To further reduce this discrepancy, one can divide the $\eta$ bins into segments of equal size in $\varphi$ and use Eq.~(\ref{eq:FinalCovFormula}) to calculate the correlation factors.
In the left pane of Fig.~\ref{fig:VarAccCorrP3Simp}, this has been done using 10 $\varphi$ segments.
While up to 40\% of the bin is inactive, detecting down to 60\% of the particles, the obtained correlation factors now agree to within a few per mill of the primary values.

The chosen number of segments in the analysis influences the accuracy of the result.
This is shown in the right pane of Fig.~\ref{fig:VarAccCorrP3Simp} where the correlation factor has been computed using 1, 5, 10, and 20 $\varphi$ segments.
Using one $\varphi$ segment produces the same result as in the right pane of Fig.~\ref{fig:VarAccP3Simp}, while choosing more segments improves the result up to having 10 segments.
The results when using 10 and 20 segments are identical.
This is true, because every adjacent pair, in $\varphi$, of acceptance values is the same and, therefore, the 20 segment version of Eq.~(\ref{eq:FinalCovFormula}) simplifies identically into the 10 segment version of that equation.
If one had, for instance, the same acceptance value for every $\varphi$ segment in an $\eta$ bin, Eq.~(\ref{eq:FinalCovFormula}) would identically simplify to Eq.~(\ref{eq:CorrectedCov}) or (\ref{eq:CorrectedVar}) depending on whether it corresponded to a covariance or a variance.
In the example in the right pane of Fig.~\ref{fig:VarAccCorrP3Simp}, 10 segments is enough to ensure segments of equal size, while also ensuring that all segments have the same detection efficiency of either 1 or 0.
This is not the case when the correlation factor is computed using 5 segments.
In that case, one (or more) segments have an average efficiency of 0.5.
This makes the 5 segment case more inaccurate because the assumption of uniform efficiency in the bin is violated.
This study shows that, while finer segmentation can produce more accurate results, there may exist a limit beyond which no further accuracy is attained.
In fact, if possible, the segmentation used in the analysis should only be fine enough to ensure that all segments have an efficiency of either 1 or 0, if acceptance is the only effect being accounted for, since this will reduce the required storage of information to perform the measurement.

\subsubsection{A Realistic Case}
\label{sec:realisticcase}
\begin{figure}[ht]
\begin{center}
\includegraphics[width=0.49\textwidth]{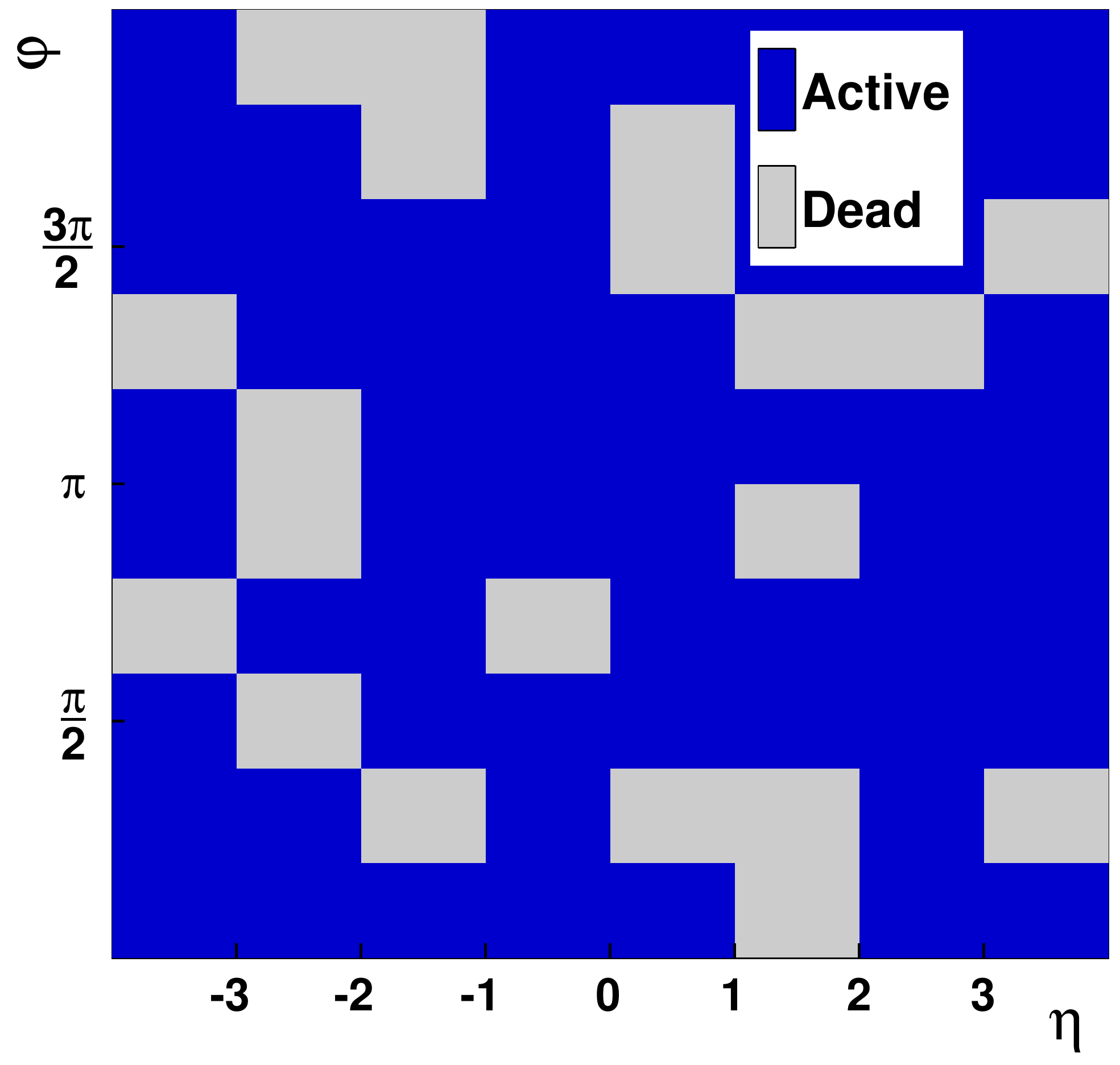} 
\includegraphics[width=0.49\textwidth]{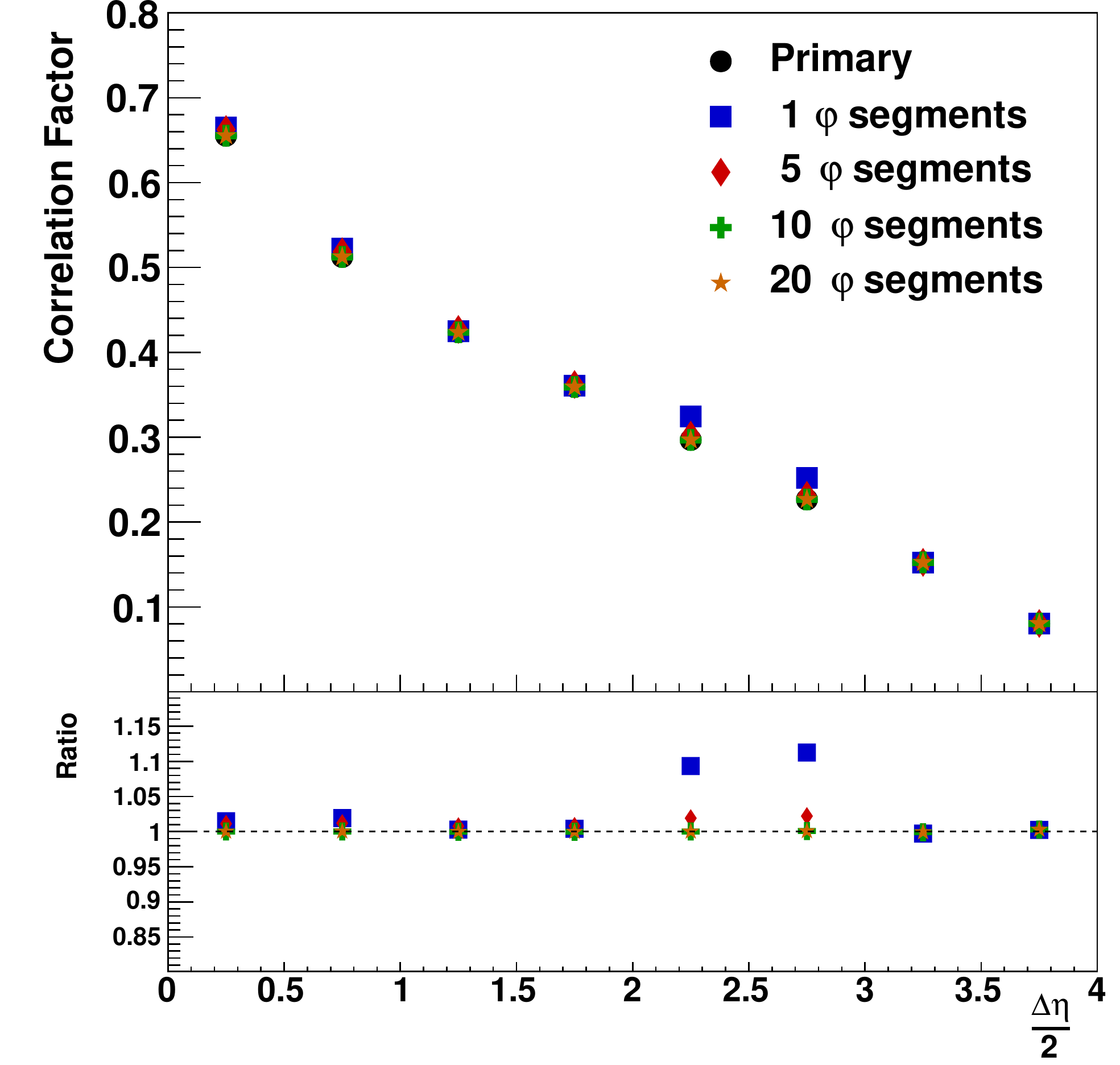}
\caption{(Color online) Left: Acceptance map with 20 randomly placed inactive regions.
The individual regions span 1 unit in $\eta$ and $\frac{2\pi}{10}$ in azimuth.
Right: The correlation factors found after accounting for randomly placed inactive regions using different $\varphi$ segmentations.
The bottom part shows the ratio of the obtained correlation factors to the primary correlation factors.}
\label{fig:AccMapRandomResult}
\end{center}
\end{figure}

The simple test shown in section~\ref{sec:simplecase} demonstrates the general effect of reduced acceptance.
Realistic detector acceptances lack that simplicity though.
To test the method more generally, 20 inactive regions were placed randomly over the analysis region.
The only restriction placed on the randomness was that there must be greater than 50\% acceptance in every $\eta$ bin to ensure that the correlation factor can be calculated using this method.
The resulting acceptance map is shown in the left pane of Fig.~\ref{fig:AccMapRandomResult}.

The right pane of Fig.~\ref{fig:AccMapRandomResult} shows the result of the analysis with different numbers of $\varphi$ segments.
When the acceptance varies in each $\eta$ bin, structure can be seen in the obtained correlation factors with no $\varphi$ segmentation that is not present in the simple case presented in section~\ref{sec:simplecase}.
Including $\varphi$ segmentation minimizes this effect.
Increasing the number of $\varphi$ segments to 10 gives the same accuracy as seen in the simple case.
Also as for the simple case, increasing the segmentation beyond 10 $\varphi$ segments in these examples does not produce a more accurate measurement.

\subsection{Efficiency}
\label{sec:efficiency}
In this section we address the case where the detection efficiency can have any value between 0 and 1.
This is in contrast to the previous cases where the detection efficiency was 1 for active regions and 0 for inactive regions.
This case is quite realistic for most detectors since perfect detection efficiency is never achieved.
A continuous efficiency gradient (in both $\varphi$ and $\eta$) is applied to the primary particles from the generator.

\subsubsection{$\varphi$ Dependent Efficiency}
\label{sec:phiefficiency}
\begin{figure}[ht]
\begin{center}
\includegraphics[width=0.49\textwidth]{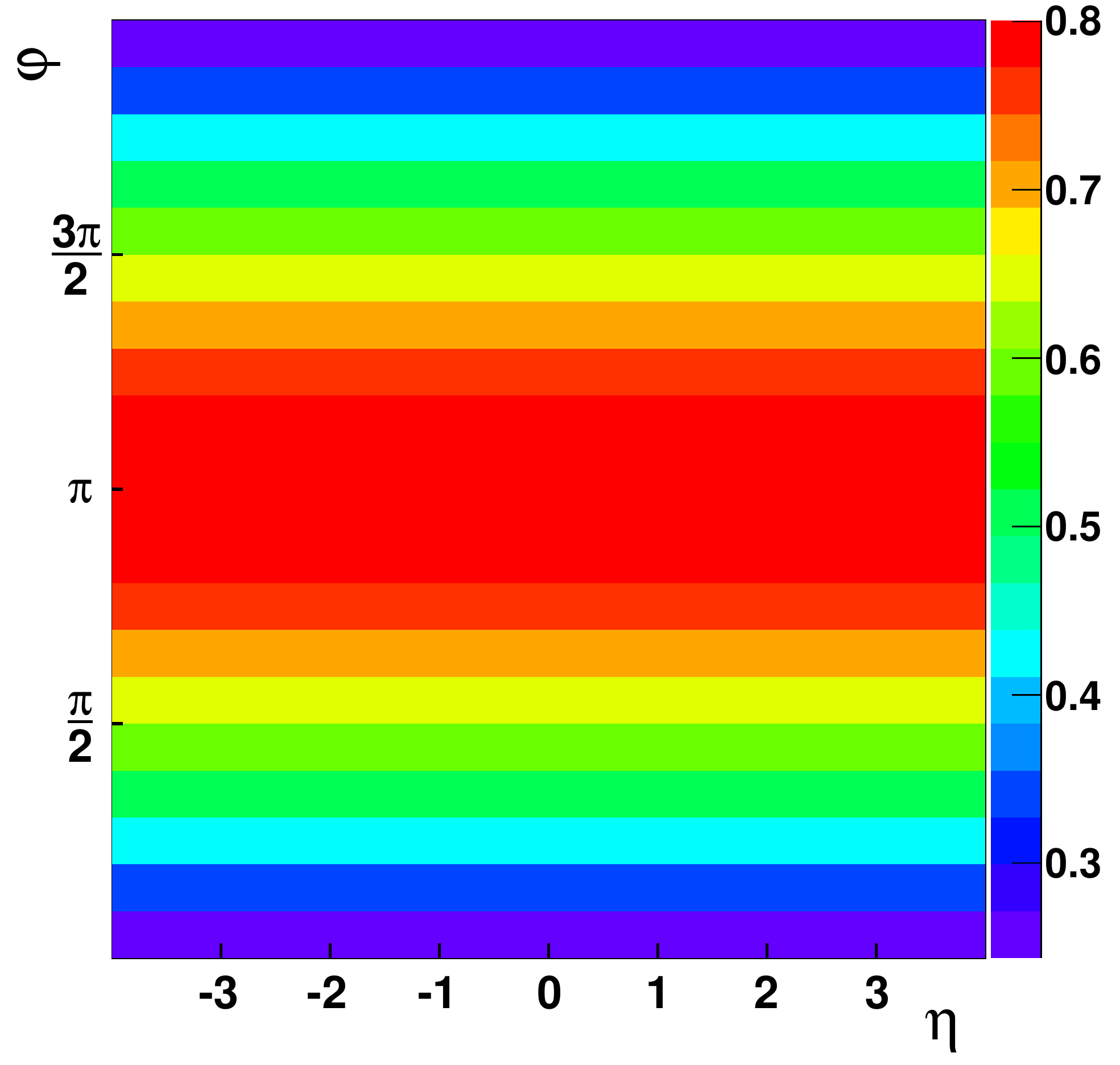}
\includegraphics[width=0.49\textwidth]{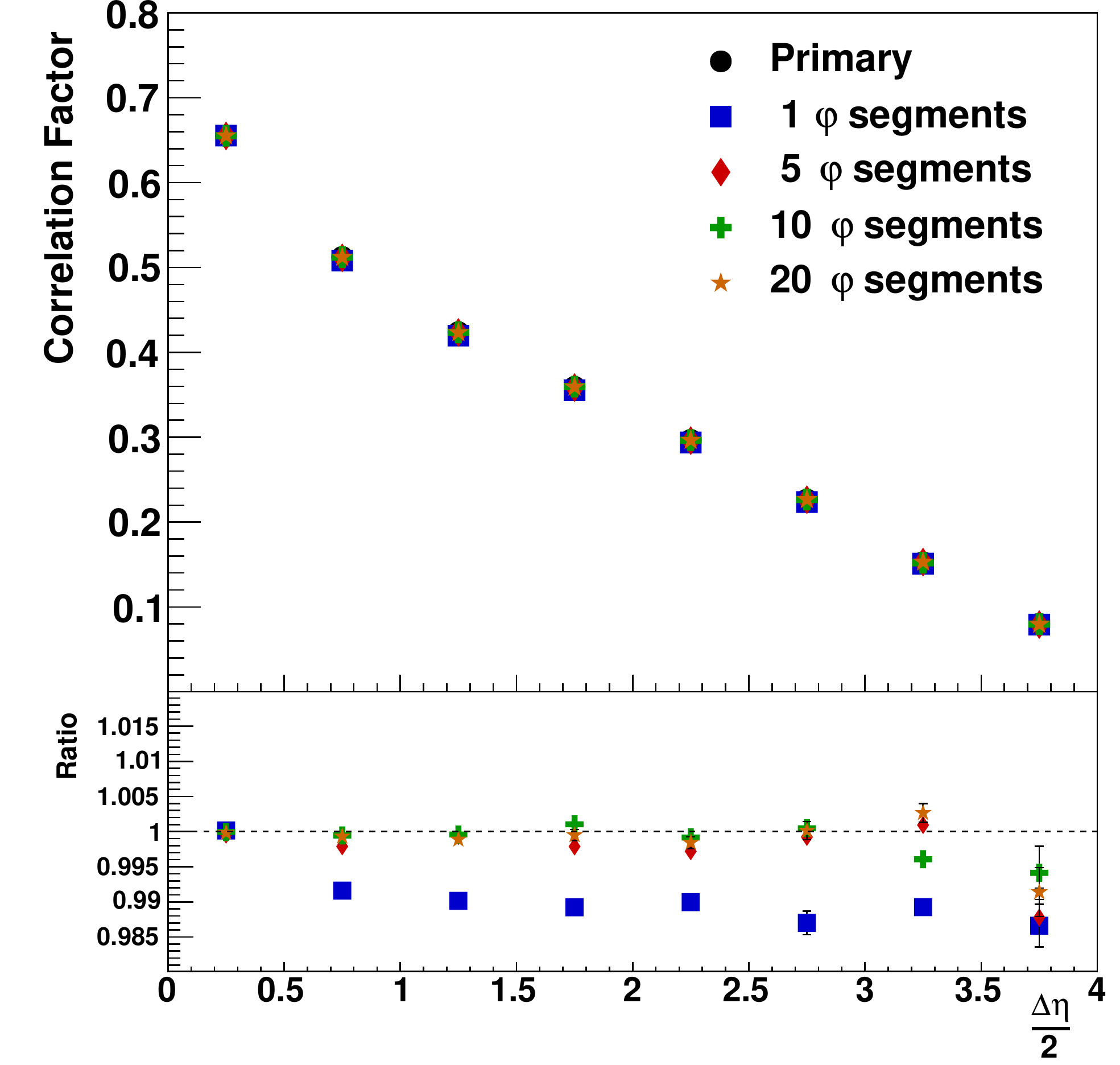}
\caption{(Color online) Left: The figure shows efficiency map from a sine function of the form $\varepsilon(\varphi)=0.6\sin(\varphi/2)+0.2$.
The efficiencies take on values in the range $0.2 \leq \varepsilon \leq 0.8$ with the azimuthal angular range of $0 \leq \varphi \leq 2\pi$.
Note that the colors indicate the average efficiency within the bins and not the values from the efficiency function itself.
Right: Resultant correlation factors obtained using Eq.~(\ref{eq:FinalCovFormula}) with efficiency values extracted from the shown efficiency map using different azimuthal segmentations.
The bottom part shows the ratio of the obtained correlation factors to the primary correlation factors.}
\label{fig:PhiEffGradMapResult}
\end{center}
\end{figure}

To study the effect of a $\varphi$ efficiency gradient, a sine function of the form $\varepsilon(\varphi)=0.6\sin(\varphi/2)+0.2$ is imposed such that the range of efficiency values is $0.2 \leq \varepsilon \leq 0.8$ for $0 \leq \varphi \leq 2\pi$. 
The resulting efficiency map is shown in the left pane of Fig.~\ref{fig:PhiEffGradMapResult}.
Note that, due to binning, the values portrayed in the figure show the average efficiency of the detection regions and not the continuous distribution which is actually imposed on the particles.

The results from applying a continuous efficiency gradient in $\varphi$ are shown in the right pane of Fig.~\ref{fig:PhiEffGradMapResult}.
In principle, the accuracy can always be improved by increasing the number of segments, because the gradient never vanishes.
In this case, one must choose the number of segments corresponding to the desired accuracy and available statistics.
In this analysis an accuracy of better than 1\% is already achieved by using 5 $\varphi$ segments. 

\subsubsection{$\eta$ Dependent Efficiency}
\label{sec:etaefficiency}
\begin{figure}[ht]
\begin{center}
\includegraphics[width=0.49\textwidth]{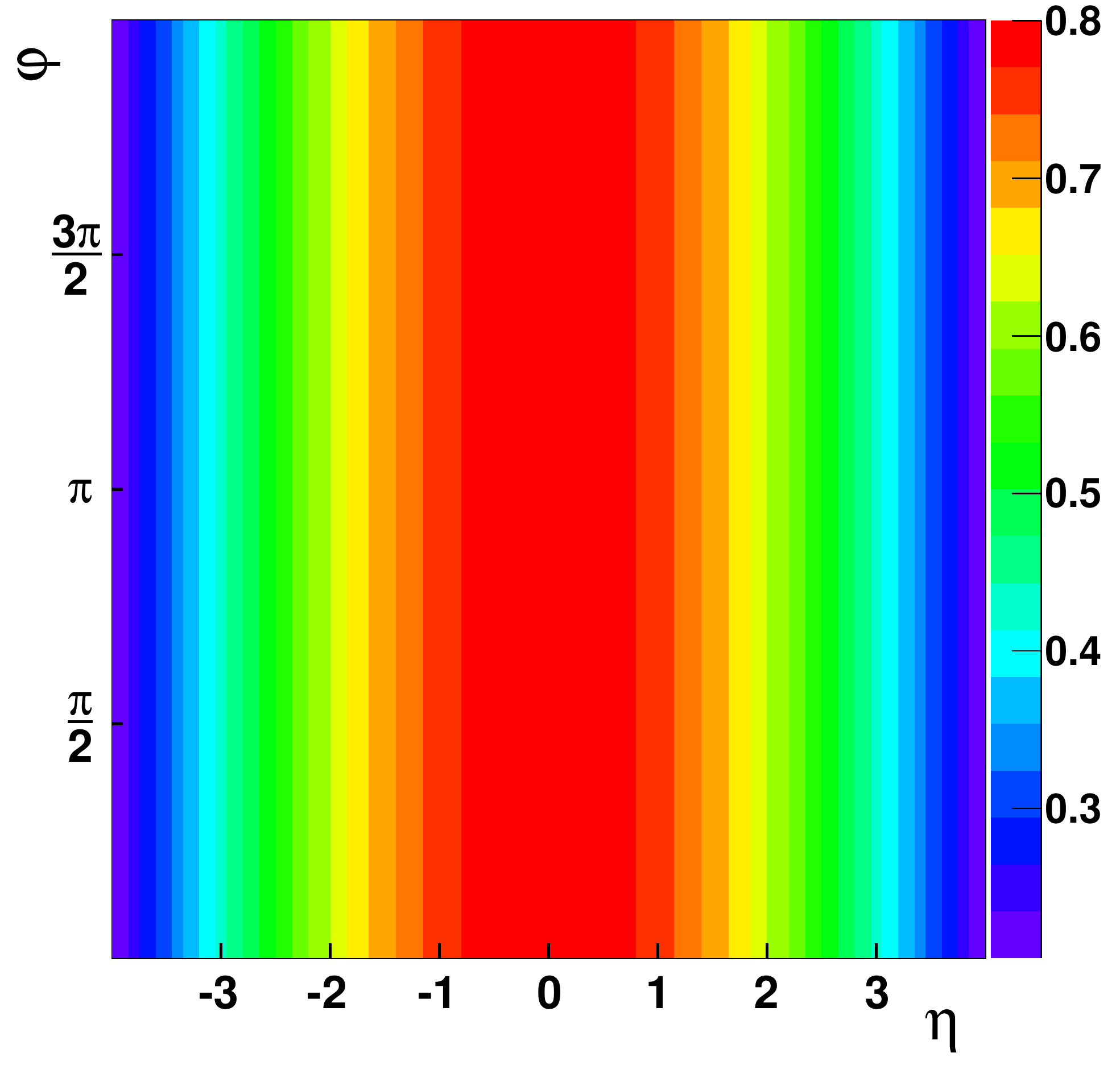}
\includegraphics[width=0.49\textwidth]{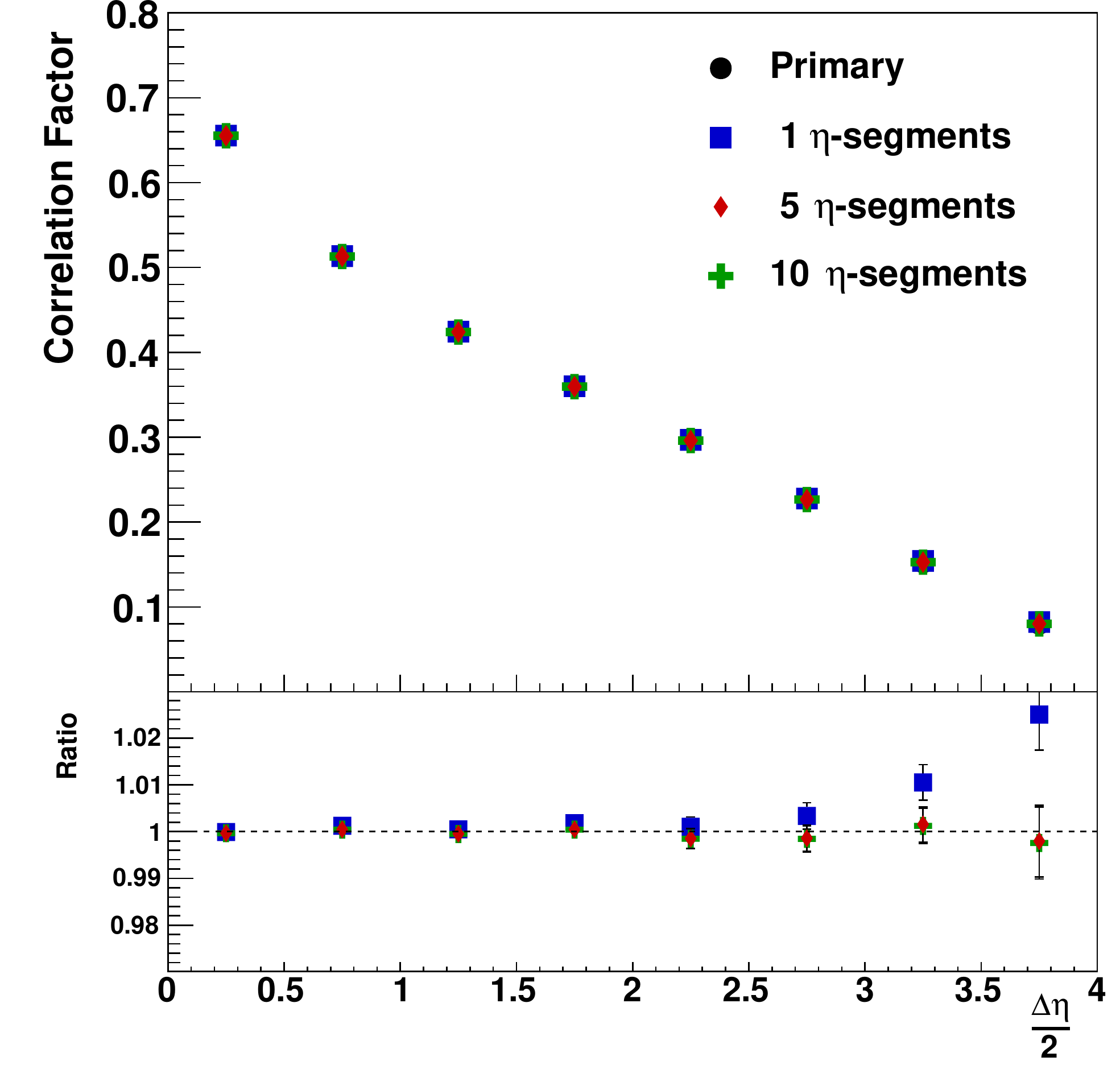}
\caption{(Color online) Left: The figure shows efficiency map from a sine function of the form $\varepsilon(\eta)=0.6\sin((\eta/4+1)\cdot\pi/2)+0.2$.
The efficiencies take on values in the range $0.2 \leq \varepsilon \leq 0.8$ with the pseudorapidity range of $-4 \leq \eta \leq 4$.
The colors indicate the average efficiency within the bins and not the values from the efficiency function itself.
Right: Resultant correlation factors obtained using Eq.~(\ref{eq:FinalCovFormulaEta}) with efficiency values extracted from the shown efficiency map using different pseudorapidity segmentations for each measured point.
The bottom part shows the ratio of the obtained correlation factors to the primary correlation factors.}
\label{fig:EtaEffGradMapResult}
\end{center}
\end{figure}

To study the effect of an $\eta$ efficiency gradient, a sine function of the form $\varepsilon(\eta)=0.6\sin((\eta/4+1)\cdot\pi/2)+0.2$ is imposed such that the range of efficiency values is again $0.2 \leq \varepsilon \leq 0.8$ for $-4 \leq \eta \leq 4$. 
The efficiency map for this gradient is shown in the left pane of Fig.~\ref{fig:EtaEffGradMapResult}.

The results from applying a continuous efficiency gradient in $\eta$ are shown in the right pane of Fig.~\ref{fig:EtaEffGradMapResult}.
Again, in principle, the accuracy can always be improved by increasing the number of segments, because the gradient never vanishes.
However, while one does see improvement increasing the $\eta$ segmentation from 1 to 5 segments per $\eta$ bin, one sees virtual no improvement continuing to 10 segments.

\subsection{Comparison between Different Tunes}
\label{sec:GenComp}
\begin{figure}[ht]
\begin{center}
\includegraphics[width=0.32\textwidth]{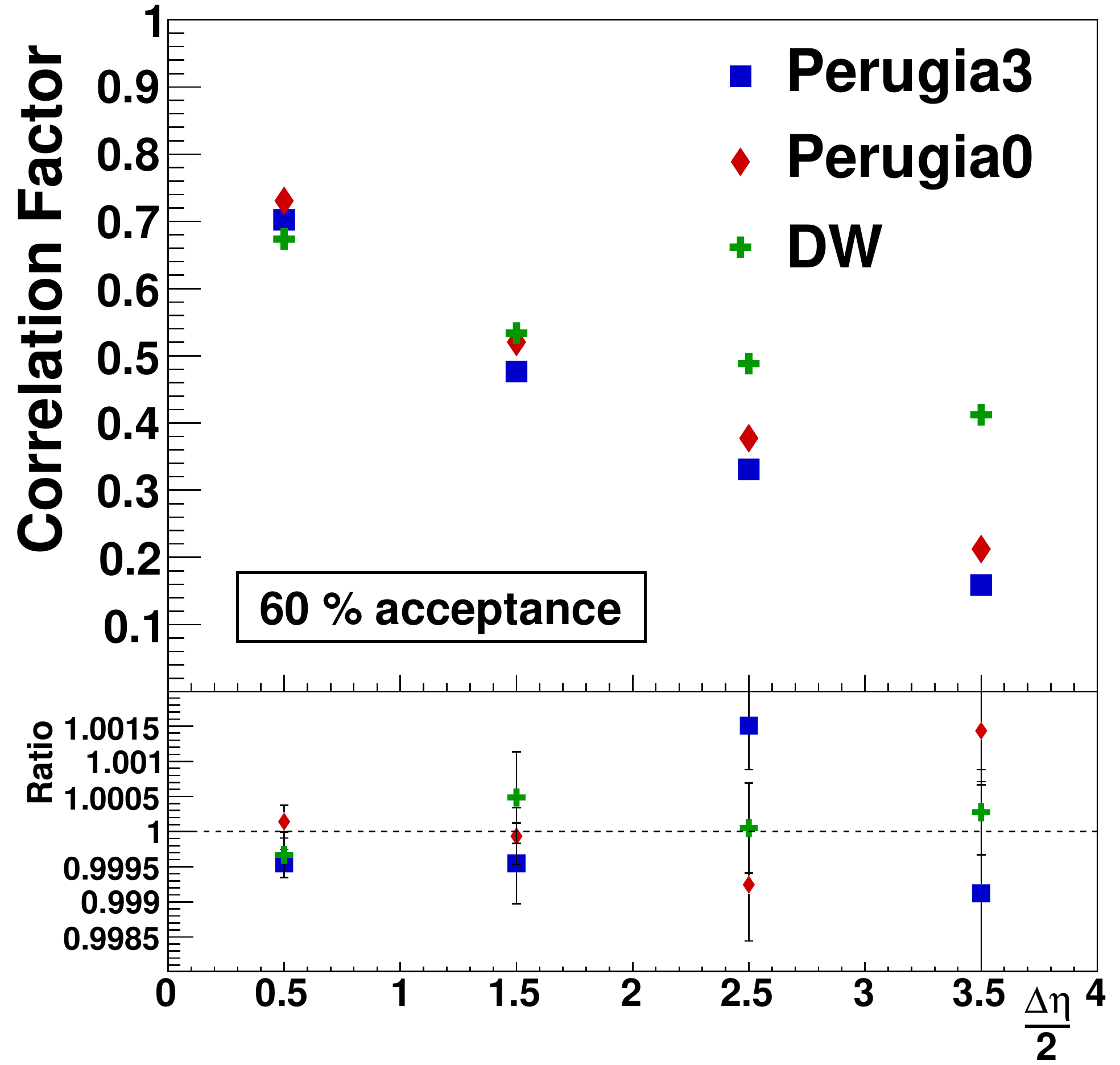}
\includegraphics[width=0.32\textwidth]{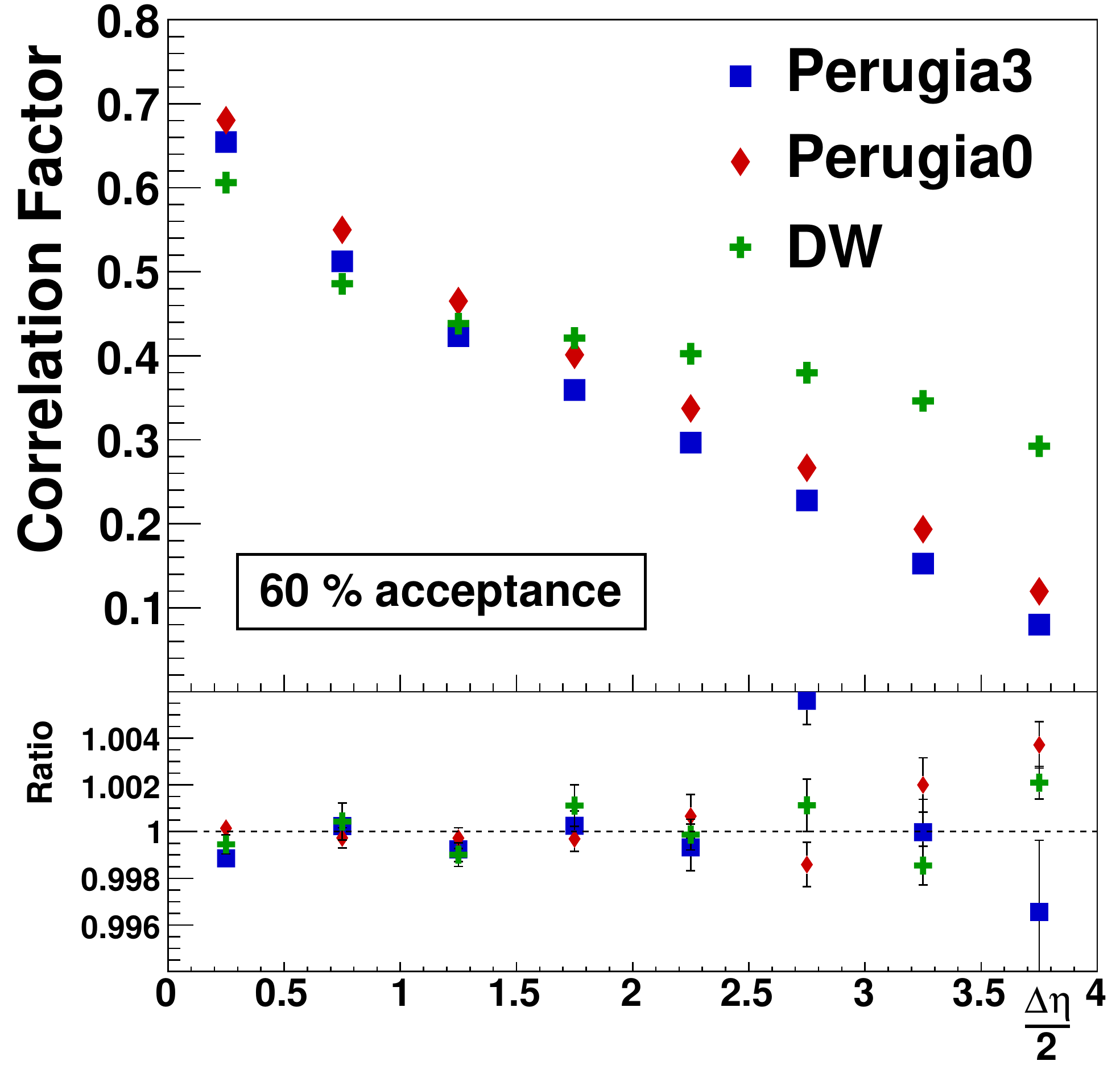}
\includegraphics[width=0.32\textwidth]{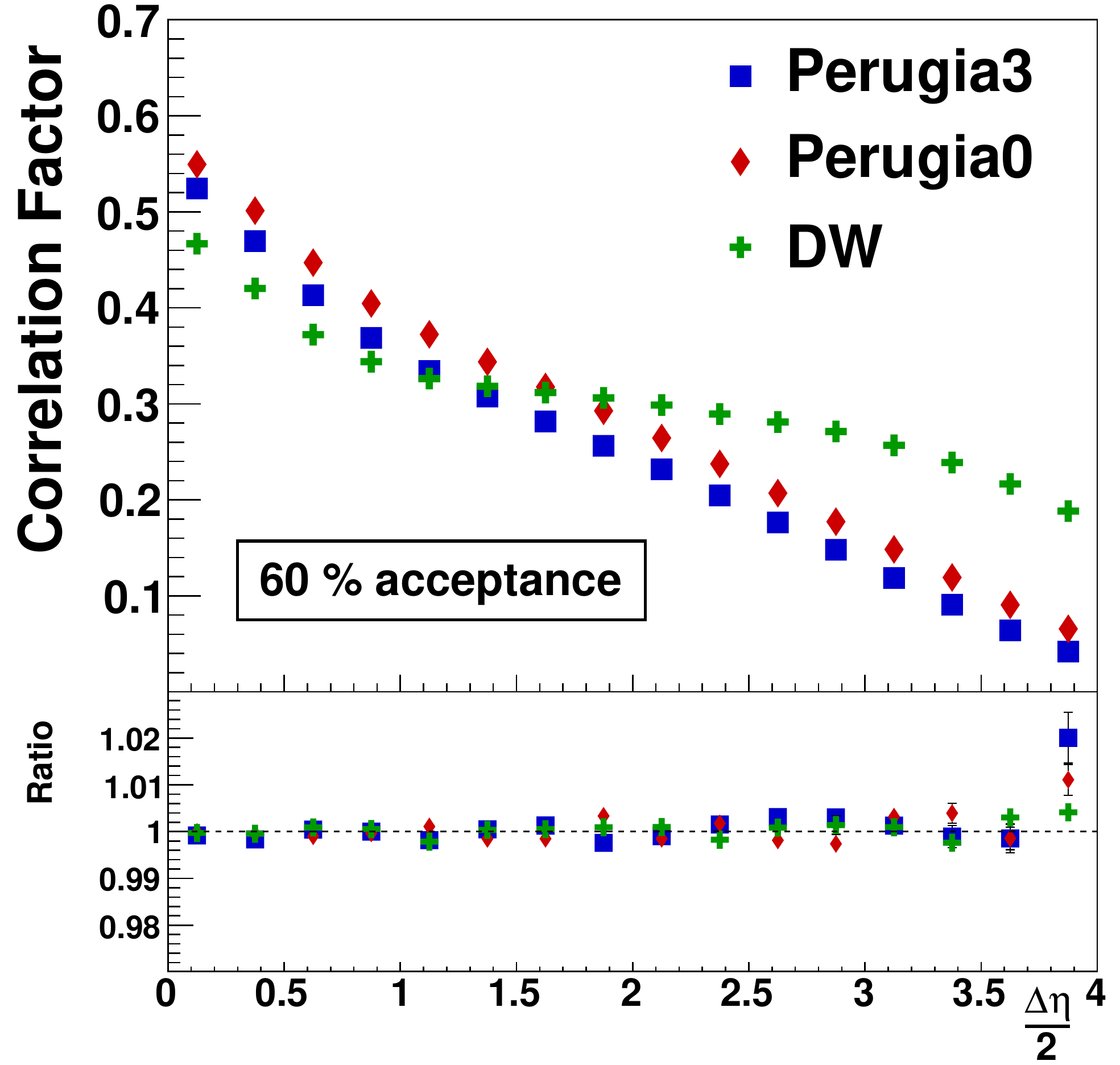}
\caption{(Color online) Resultant correlation factors obtained using different tunes of Pythia and three different $\eta$ bin widths ($\Delta_{\textrm{bin}} = 1, 0.5,\textrm{ and } 0.25$).
Data from each tune are subjected to the same geometrical acceptance (60\%).
The bottom part of the figures shows the ratio of the obtained correlation factor to the primary value for that tune (shown in Fig.~\ref{fig:PrimComp}).
The figures show no significant dependence of the deviations on the tune.}
\label{fig:VariedGenerator}
\end{center}
\end{figure}

The need for such accuracy achieved with this method can be shown by looking at the results from the different generators.
Figure~\ref{fig:VariedGenerator} shows the results using the same particles that produced the curves in Fig.~\ref{fig:PrimComp} with the 60\% simple acceptance configuration in each $\eta$ bin applied.
The bins were divided into 10 azimuthal segments and Eq.~(\ref{eq:FinalCovFormula}) was used to obtain the results.
The method has been applied with no simulation input and only the knowledge of the acceptance for all results.
The results show no particularly different behavior for the discrepancies from the true values for any specific generator (tune).
For the vast majority of points, for all bin widths, the accuracy of the obtained values is within 1\%.

Figure~\ref{fig:VariedGenerator} reproduces the curves in Fig.~\ref{fig:PrimComp}.
For small $\Delta\eta$ (which many detectors possess), one must achieve high accuracy and precision to distinguish between different tunes and, consequently, the relative strengths of the underlying physical processes.
The methods presented here allow one to make correlation measurements with high enough accuracy and precision to achieve this goal.

\section{Conclusions}
\label{sec:conclusion}
The effect of reduced acceptance and imperfect detection efficiency on forward-backward correlations is derived using a statistical approach.
No assumptions about the distribution of primary particles were made and, therefore, the derived results are valid for physical data as well as the simulated data studied here. 
Furthermore a framework to evaluate the effect of detector acceptance and efficiency on any order multiplicity correlation has been established.
If the acceptance and the efficiency are well determined, the method can evaluate forward-backward correlations very accurately depending on the capabilities of the detector (segmentation) as long as the inactive regions are smaller than 50\% in all $\eta$ regions.
Considerations must be made concerning the desired segmentation used in the analysis.
The number of segments should be large enough to ensure nearly constant efficiency within the segments while balancing against the storage required for recording the necessary information for the analysis. 

The presented method allows one to achieve high accuracy for computing multiplicity correlations necessary to distinguish between the underlying processes governing particle production in the collision.
The framework could be further used to investigate higher order multiplicity correlations that could put additional constraints on models.
To further gain the power to distinguish between the underlying processes, one must allow for these correlation measurements to be performed accurately with large $\Delta\eta$.
This often requires using detectors which have little ability to reject secondary particles (which this paper has not investigated).
Extending this framework to deal with this effect would provide a powerful tool in the analysis of correlations over wide $\eta$ ranges.

\section*{Acknowledgments}
We would like to thank the Danish National Research Foundation (DNRF), the Danish Natural Science Research Council (FNU), and the Villum Foundation for their financial support of this research.

\bibliography{bibliography}

\begin{thebibliography}{8}%
\makeatletter
\providecommand \@ifxundefined [1]{%
 \@ifx{#1\undefined}
}%
\providecommand \@ifnum [1]{%
 \ifnum #1\expandafter \@firstoftwo
 \else \expandafter \@secondoftwo
 \fi
}%
\providecommand \@ifx [1]{%
 \ifx #1\expandafter \@firstoftwo
 \else \expandafter \@secondoftwo
 \fi
}%
\providecommand \natexlab [1]{#1}%
\providecommand \enquote  [1]{``#1''}%
\providecommand \bibnamefont  [1]{#1}%
\providecommand \bibfnamefont [1]{#1}%
\providecommand \citenamefont [1]{#1}%
\providecommand \href@noop [0]{\@secondoftwo}%
\providecommand \href [0]{\begingroup \@sanitize@url \@href}%
\providecommand \@href[1]{\@@startlink{#1}\@@href}%
\providecommand \@@href[1]{\endgroup#1\@@endlink}%
\providecommand \@sanitize@url [0]{\catcode `\\12\catcode `\$12\catcode
  `\&12\catcode `\#12\catcode `\^12\catcode `\_12\catcode `\%12\relax}%
\providecommand \@@startlink[1]{}%
\providecommand \@@endlink[0]{}%
\providecommand \url  [0]{\begingroup\@sanitize@url \@url }%
\providecommand \@url [1]{\endgroup\@href {#1}{\urlprefix }}%
\providecommand \urlprefix  [0]{URL }%
\providecommand \Eprint [0]{\href }%
\providecommand \doibase [0]{http://dx.doi.org/}%
\providecommand \selectlanguage [0]{\@gobble}%
\providecommand \bibinfo  [0]{\@secondoftwo}%
\providecommand \bibfield  [0]{\@secondoftwo}%
\providecommand \translation [1]{[#1]}%
\providecommand \BibitemOpen [0]{}%
\providecommand \bibitemStop [0]{}%
\providecommand \bibitemNoStop [0]{.\EOS\space}%
\providecommand \EOS [0]{\spacefactor3000\relax}%
\providecommand \BibitemShut  [1]{\csname bibitem#1\endcsname}%
\let\auto@bib@innerbib\@empty
%</preamble>
\bibitem [{\citenamefont {Sjostrand}\ and\ \citenamefont {van
  Zijl}(1987)}]{Sjostrand:1987su}%
  \BibitemOpen
  \bibfield  {author} {\bibinfo {author} {\bibfnamefont {T.}~\bibnamefont
  {Sjostrand}}\ and\ \bibinfo {author} {\bibfnamefont {M.}~\bibnamefont {van
  Zijl}},\ }\href {\doibase 10.1103/PhysRevD.36.2019} {\bibfield  {journal}
  {\bibinfo  {journal} {Phys.Rev.}\ }\textbf {\bibinfo {volume} {D36}},\
  \bibinfo {pages} {2019} (\bibinfo {year} {1987})}\BibitemShut {NoStop}%
%%CITATION = PHRVA,D36,2019;%%
\bibitem [{\citenamefont {Hwa}\ and\ \citenamefont {Yang}(2007)}]{Hwa:2007sq}%
  \BibitemOpen
  \bibfield  {author} {\bibinfo {author} {\bibfnamefont {R.~C.}\ \bibnamefont
  {Hwa}}\ and\ \bibinfo {author} {\bibfnamefont {C.}~\bibnamefont {Yang}},\
  }\href@noop {} {\  (\bibinfo {year} {2007})},\ \Eprint
  {http://arxiv.org/abs/0705.3073} {arXiv:0705.3073 [nucl-th]} \BibitemShut
  {NoStop}%
%%CITATION = ARXIV:0705.3073;%%
\bibitem [{\citenamefont {Wraight}\ and\ \citenamefont
  {Skands}(2011)}]{Wraight:2011ej}%
  \BibitemOpen
  \bibfield  {author} {\bibinfo {author} {\bibfnamefont {K.}~\bibnamefont
  {Wraight}}\ and\ \bibinfo {author} {\bibfnamefont {P.}~\bibnamefont
  {Skands}},\ }\href {\doibase 10.1140/epjc/s10052-011-1628-z} {\bibfield
  {journal} {\bibinfo  {journal} {Eur.Phys.J.}\ }\textbf {\bibinfo {volume}
  {C71}},\ \bibinfo {pages} {1628} (\bibinfo {year} {2011})},\ \Eprint
  {http://arxiv.org/abs/1101.5215} {arXiv:1101.5215 [hep-ph]} \BibitemShut
  {NoStop}%
%%CITATION = ARXIV:1101.5215;%%
\bibitem [{\citenamefont {Skands}(2009)}]{Skands:2009zm}%
  \BibitemOpen
  \bibfield  {author} {\bibinfo {author} {\bibfnamefont {P.~Z.}\ \bibnamefont
  {Skands}},\ }\href@noop {} {\  (\bibinfo {year} {2009})},\ \Eprint
  {http://arxiv.org/abs/0905.3418} {arXiv:0905.3418 [hep-ph]} \BibitemShut
  {NoStop}%
%%CITATION = ARXIV:0905.3418;%%
\bibitem [{\citenamefont {S\o{}gaard}(2012)}]{soegaardPhD}%
  \BibitemOpen
  \bibfield  {author} {\bibinfo {author} {\bibfnamefont {C.}~\bibnamefont
  {S\o{}gaard}},\ }\emph {\bibinfo {title} {Measurement of Forward-Backward
  Charged Particle Correlations with ALICE}},\ \href
  {http://www.nbi.dk/~soegaard/SoegaardPhDThesis.pdf} {Ph.D. thesis},\ \bibinfo
   {school} {Niels Bohr Institute, University of Copenhagen} (\bibinfo {year}
  {2012})\BibitemShut {NoStop}%
\bibitem [{\citenamefont {Ravan}\ \emph {et~al.}(2014)\citenamefont {Ravan},
  \citenamefont {Pujahari}, \citenamefont {Prasad},\ and\ \citenamefont
  {Pruneau}}]{Ravan:2013lwa}%
  \BibitemOpen
  \bibfield  {author} {\bibinfo {author} {\bibfnamefont {S.}~\bibnamefont
  {Ravan}}, \bibinfo {author} {\bibfnamefont {P.}~\bibnamefont {Pujahari}},
  \bibinfo {author} {\bibfnamefont {S.}~\bibnamefont {Prasad}}, \ and\ \bibinfo
  {author} {\bibfnamefont {C.~A.}\ \bibnamefont {Pruneau}},\ }\href {\doibase
  10.1103/PhysRevC.89.024906} {\bibfield  {journal} {\bibinfo  {journal}
  {Phys.Rev.}\ }\textbf {\bibinfo {volume} {C89}},\ \bibinfo {pages} {024906}
  (\bibinfo {year} {2014})},\ \Eprint {http://arxiv.org/abs/1311.3915}
  {arXiv:1311.3915 [nucl-ex]} \BibitemShut {NoStop}%
%%CITATION = ARXIV:1311.3915;%%
\bibitem [{\citenamefont {Sjostrand}\ \emph {et~al.}(2006)\citenamefont
  {Sjostrand}, \citenamefont {Mrenna},\ and\ \citenamefont
  {Skands}}]{Sjostrand:2006za}%
  \BibitemOpen
  \bibfield  {author} {\bibinfo {author} {\bibfnamefont {T.}~\bibnamefont
  {Sjostrand}}, \bibinfo {author} {\bibfnamefont {S.}~\bibnamefont {Mrenna}}, \
  and\ \bibinfo {author} {\bibfnamefont {P.~Z.}\ \bibnamefont {Skands}},\
  }\href {\doibase 10.1088/1126-6708/2006/05/026} {\bibfield  {journal}
  {\bibinfo  {journal} {JHEP}\ }\textbf {\bibinfo {volume} {0605}},\ \bibinfo
  {pages} {026} (\bibinfo {year} {2006})},\ \Eprint
  {http://arxiv.org/abs/hep-ph/0603175} {arXiv:hep-ph/0603175 [hep-ph]}
  \BibitemShut {NoStop}%
%%CITATION = HEP-PH/0603175;%%
\bibitem [{\citenamefont {Skands}(2010)}]{Skands:2010ak}%
  \BibitemOpen
  \bibfield  {author} {\bibinfo {author} {\bibfnamefont {P.~Z.}\ \bibnamefont
  {Skands}},\ }\href {\doibase 10.1103/PhysRevD.82.074018} {\bibfield
  {journal} {\bibinfo  {journal} {Phys.Rev.}\ }\textbf {\bibinfo {volume}
  {D82}},\ \bibinfo {pages} {074018} (\bibinfo {year} {2010})},\ \Eprint
  {http://arxiv.org/abs/1005.3457} {arXiv:1005.3457 [hep-ph]} \BibitemShut
  {NoStop}%
%%CITATION = ARXIV:1005.3457;%%
\end{thebibliography}%

\end{document}